\documentclass[conference]{IEEEtran}
\IEEEoverridecommandlockouts
\usepackage{acronym}

\usepackage{color}

\usepackage{pgfplots}
\usepackage{tikz}
\usetikzlibrary{calc}
\makeatletter
\newcommand{\gettikzxy}[3]{%
  \tikz@scan@one@point\pgfutil@firstofone#1\relax
  \edef#2{\the\pgf@x}%
  \edef#3{\the\pgf@y}%
}
\usetikzlibrary{spy,backgrounds}
\usepackage{mathrsfs}
\usepackage{booktabs} 
\usepackage{amsmath}
\usepackage{graphicx} 
\usepackage{epstopdf}
\usepackage{amssymb}
\usepackage{amsfonts}
\usepackage{amsthm}
\usepackage{cite}
\usepackage{bm,comment}
\usepackage{algorithm}
\usepackage{algpseudocode}

\usepackage{subeqnarray}
\usepackage{subfigure}
\usepackage{multicol}
\usepackage{multirow}
\usepackage{diagbox}
\usepackage{stfloats}
\usepackage{float}
\usepackage{color} 
\usepackage{cases}
\usepackage{lipsum}

\usepackage[bookmarks,colorlinks]{hyperref} 
\hypersetup{colorlinks,citecolor= red,filecolor= blue,linkcolor= blue,urlcolor=blue}
\allowdisplaybreaks
\begin{document}
\setlength{\textfloatsep}{4pt}

\bstctlcite{IEEEexample:BSTcontrol}

\title{Integrated Positioning and Communication for Cooperative Multi-LEO Uplink Communications: A Dual-Timescale Kalman Filter-Aided Approach}

\author{Ali Hanif, \IEEEmembership{Student Member, IEEE}, Yuchen Zhang, \IEEEmembership{Member, IEEE}, Pinjun Zheng, \IEEEmembership{Member, IEEE},\\ and Tareq Y. Al-Naffouri, \IEEEmembership{Fellow, IEEE}
\thanks{A. Hanif, Y. Zhang, and  T. Y. Al-Naffouri are with the Department of Electrical and Computer Engineering, CEMSE Division, King Abdullah University of Science and Technology  (KAUST), Saudi Arabia (Email: \{ali.hanif, yuchen.zhang, tareq.alnaffouri\}@kaust.edu.sa).  P. Zheng is with the University of British Columbia, Canada
 (Email: pinjun.zheng@ubc.ca).}}

\maketitle

\begin{abstract}
Low Earth orbit (LEO) satellites are a crucial component of the future non-terrestrial networks (NTN) due to lower latency, robust signal strengths, shorter revisit times, and dense constellations. However, acquiring reliable channel state information (CSI) in LEO satellite communication remains challenging owing to severe signal attenuation over long propagation distances and short coherence times. Despite these challenges, LEO channels benefit from pronounced line-of-sight dominance and geometric properties inherently tied to positioning information.
In this work, we propose an integrated positioning and communication (IPAC) framework for multi-LEO satellite networks to address the unique challenges posed by LEO channels. Specifically, we leverage in-the-loop LEO positioning to exploit users' position information for improving uplink CSI acquisition. To overcome the link-budget limitations of single-satellite systems, cooperative multi-LEO uplink data detection is adopted. By exploiting the different coherent timescales of position-related parameters and random channel gains, we develop a dual-timescale Kalman filter-based IPAC framework: an unscented Kalman filter (UKF) for tracking users' position and velocity in the large-timescale, and a Kalman filter that leverages the position information obtained in the large-timescale for improved data-aided uplink channel estimation in the small-timescale. Finally, the two tasks of channel estimation and cooperative data detection are jointly addressed through the expectation maximization (EM) algorithm. Numerical results demonstrate that the proposed IPAC approach outperforms the conventional baseline in terms of channel estimation accuracy and communication performance.
\end{abstract}


\section{INTRODUCTION}
\IEEEPARstart{N}{}on-terrestrial networks (NTN), encompassing airborne and spaceborne platforms such as high-altitude platforms (HAPs) and satellites, are anticipated to play a crucial role in 6G by providing seamless global coverage, particularly in underserved and remote areas where deploying terrestrial communication infrastructure is not economically feasible \cite{heo2023mimo,andrews20246}. Moreover, NTNs can ensure emergency communication services during natural disasters, such as earthquakes and floods, when terrestrial networks are disrupted or unavailable. Among satellite-based NTN platforms, LEO satellites being much closer to the Earth with dense constellations, are of particular interest. Compared to other space-borne platforms, they offer lower latency, stronger signal, shorter revisit time \cite{liu2021leo}, making them a promising candidate for positioning and communication applications \cite{heo2023mimo,ferre2022leo,zheng2024leo,zhang2025positioning}. However, reliable channel state information (CSI) acquisition, which is of vital importance for high-speed communication, is very challenging in LEO satellite communication (SATCOM) due to the limited link budgets and fast-varying channels \cite{heo2023mimo,li2023channel,10750262,12}. While Doppler compensation techniques at the receiver can alleviate channel aging effects caused by the high mobility of LEO satellites \cite{10750262,li2023channel,930095}, the inherently long propagation distance, compared to terrestrial networks, leads to severe signal power attenuation. This challenge becomes even more pronounced in the uplink phase, where the user terminal (UT) operates under stringent transmit power constraints. To mitigate the limited link budget inherent in single-satellite systems, multi-satellite cooperative communication \cite{zhang2025positioning,10851844,zhang2025cooperative,9939157,zhang2024multi} has emerged as a promising solution. In this approach, received signals from multiple LEO satellites are jointly processed at a central satellite node, thereby enhancing the effective signal strength.

Channel estimation schemes for LEO satellite communication can be broadly categorized into conventional pilot-based \cite{li2023channel,arti2015channel,arti2016channel,gappmair2014new} and data-aided schemes \cite{lin2024joint,11}. Pilot-based schemes rely on periodically inserted pilot signals, known to the receiver, and decouple the channel estimation from data detection. Unlike conventional pilot-based channel estimation, data-aided channel estimation leverages both pilot and data symbols in the tracking process, thereby enhancing spectral efficiency by using fewer pilots.
Kalman filter (KF) has been employed for data-aided channel tracking in both terrestrial and non-terrestrial communication scenarios \cite{7,8,9,10,11,lin2024joint}.  There is limited work reported in the literature employing Kalman filtering for channel tracking in NTNs. In \cite{lin2024joint}, joint Doppler and channel angle tracking is performed by incorporating the physical orbit characteristics of LEO satellites and UTs into the state evolution and measurement models of an extended KF. The work in \cite{11} compared block-based and symbol-based Kalman channel tracking schemes for LEO SATCOM. However, these works do not exploit UT's positioning information, which can be exploited to improve and simplify the channel tracking process. Additionally, they assume orthogonal subcarrier allocation across multiple UTs while neglecting the effect of inter-user interference (IUI). Lastly, these works still suffer from limited link budgets, and channel estimation remains a core challenge to achieve reliable high-throughput communication.

By leveraging the distinct features of LEO satellite channels, notably their strong line-of-sight (LoS) dominance and geometric attributes \cite{arapoglou2010mimo}, the inherent CSI acquisition challenges can be mitigated. These distinct features of LEO channels are inherently tied to position information, indicating that strategic use of UT positioning could help overcome channel estimation challenges in LEO satellite communications \cite{ma2024integrated}. Exploiting position information has been shown to improve communication performance in terrestrial networks \cite{di2014location,wymeersch20185g,muns2019beam,collaborative,wei2022toward,zhang2025positioning}. For instance, \cite{muns2019beam} shows how position information can aid beam alignment in autonomous vehicular communications. Similarly, \cite{di2014location} provides an overview of how position awareness can be leveraged across different layers of the protocol stack, while \cite{wei2022toward} discusses the importance of position information in the context of secure communications.

Position information in the aforementioned works is typically obtained using global navigation satellite system (GNSS). However, reliance on GNSS poses challenges, including high latency and coverage holes in environments such as urban canyons \cite{wymeersch20185g}. Luckily, LEO satellite positioning provides a promising alternative way to complement GNSS due to stronger signals, lower latency, enhanced visibility, and greater frequency diversity \cite{ferre2022leo,zheng2024leo,zhang2025positioning,kassas2024ad}. 
Existing research on integrating positioning and communication for LEO satellites focuses on trade-offs between positioning and communication \cite{you2024integrated} while overlooking the mutual benefits. In the same context, we propose to exploit the position information of UTs, obtained using in-the-loop LEO positioning rather than relying on external sources like GNSS, to improve CSI acquisition in LEO SATCOM scenarios, a concept we refer to as integrated positioning and communication (IPAC) in this work. Furthermore, to enable the use of in-the-loop LEO positioning, we propose to employ a dual-timescale approach that allows leveraging position information from the large-timescale while estimating the channel gains on the small-timescale. Detailed elaboration of dual-timescale frame structure and its implications are presented in Section \ref{frame}.

In this work, we investigate a multi-LEO satellite system, where each satellite is equipped with a uniform planar array (UPA) and simultaneously provides downlink positioning and uplink communication services to multiple single-antenna terrestrial UTs. Our objective is to address the key challenges inherent to this scenario: mitigating the link-budget limitations of single-satellite systems through cooperative multi-LEO data detection, employing KF-enabled data-aided channel estimation to improve spectral efficiency, and leveraging UT's position information to enhance channel estimation. By exploiting the different coherent timescales of position-related parameters and random channel gains, we propose a dual-timescale KF-based IPAC framework: an unscented Kalman filter (UKF) for tracking users' position and velocity in the large-timescale, and a KF that leverages the position information obtained in the large-timescale for improved data-aided uplink channel estimation in the small-timescale. Finally, the expectation maximization (EM) algorithm is employed to perform uplink channel estimation and cooperative multi-LEO data detection jointly.

The main contributions of this work are summarized as follows:
\begin{itemize}
    \item We propose a novel IPAC framework for improved cooperative multi-LEO uplink communication performance. The framework exploits position information obtained during the downlink positioning phase to facilitate muti-LEO cooperative joint uplink channel estimation and data detection (hereafter referred to as JUDE), thereby enhancing both channel estimation accuracy and communication performance.
    \item In the large-timescale, a UKF is employed for tracking UTs' position and velocity. In the small-timescale, a KF-based JUDE approach effectively estimates and tracks LEO channels with improved accuracy while reducing pilot overhead by exploiting both pilot and data symbols, thereby enhancing spectral efficiency. To overcome the limited link budget inherent in single-satellite systems, multi-satellite cooperative uplink communication is employed to improve the data detection performance in the presence of IUI.
    \item Numerical results demonstrate that the proposed IPAC approach outperforms the conventional pilot-based channel-estimation-then-data-detection baseline in terms of channel estimation accuracy and communication performance.
\end{itemize}

\textbf{Organization:} The system model is presented in Section II. Section III provides a detailed coverage of the proposed framework, while key results are presented and discussed in Section IV. The paper is concluded in Section V.

\textbf{Notations:} Bold lower case letters (${\bf x}$) denote vectors and upper case letters (${\bf X}$) represent matrices. The identity matrix of dimension $N\times N$ is denoted by ${\bf I}_N$ and $\mathbf{0}_N$ represents a length $N$ vector of zeros. Transpose and conjugate transposition of a matrix are denoted by $(\cdot)^T$ and $(\cdot)^H$, respectively. The conjugate is denoted by $(\cdot)^*$, and $E{(\cdot)}$ is the expectation operator. For a vector $\mathbf{a}$, its $\ell_2$-norm is expressed as $\|\mathbf{a}\|$. Complex numbers are handled by $\Re\{a\}$ and $\Im\{a\}$ for their real and imaginary components. A diagonal matrix constructed from the elements of vector $\mathbf{x}$ is written as $\operatorname{diag}(\mathbf{x})$. A block-diagonal matrix constructed from matrices $\mathbf{A}_1, \ldots, \mathbf{A}_N$ is written as $\operatorname{blkdiag}$ $\left(\mathbf{A}_1, \ldots, \mathbf{A}_N\right)$. The Kronecker product is symbolized by $\otimes$. For statistical distributions, $\mathcal{C N}(\boldsymbol{\mu}, \mathbf{C})$ represents a circularly symmetric complex Gaussian distribution characterized by mean $\boldsymbol{\mu}$ and covariance matrix $\mathbf{C}$. 

\section{SYSTEM MODEL}
\begin{figure}
    \centering
    \includegraphics[width=0.5\textwidth]{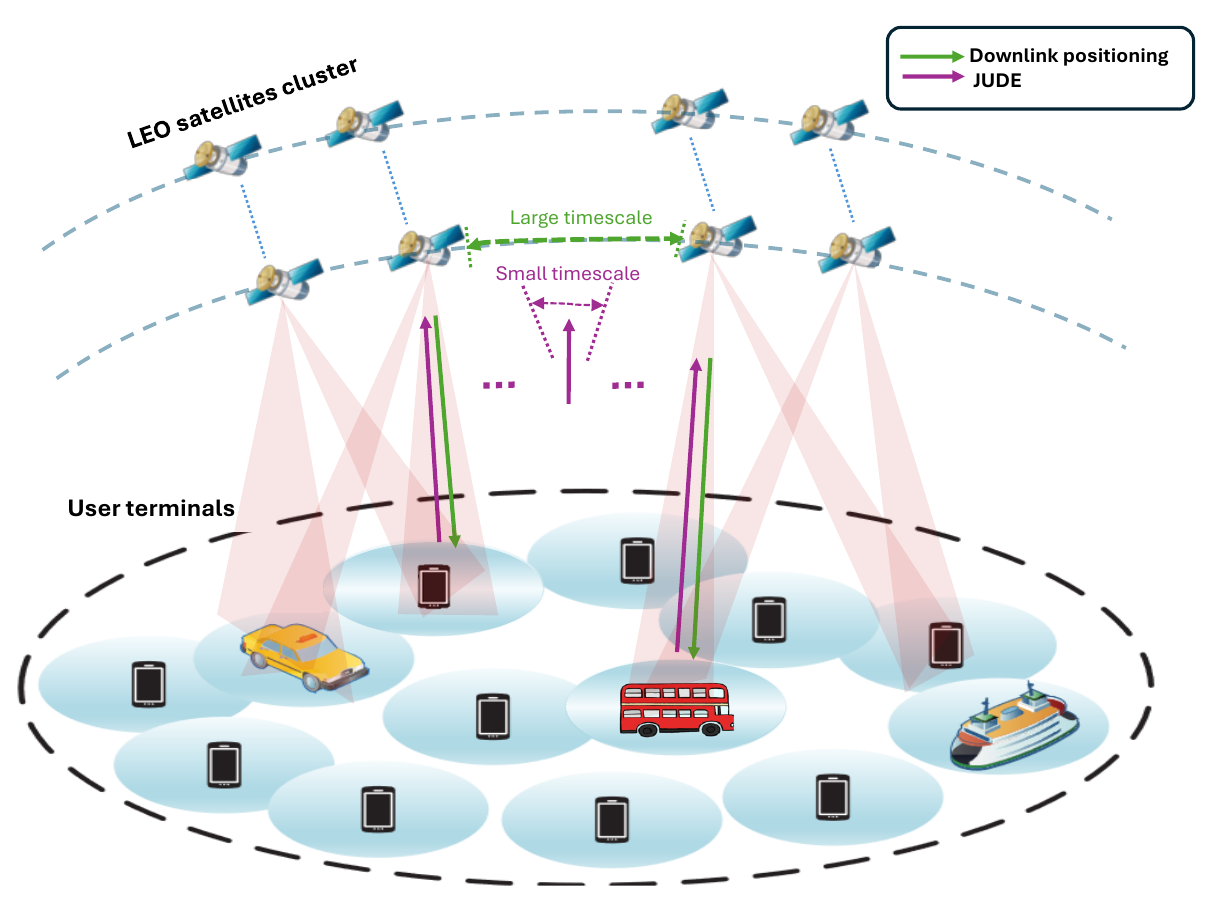}
    \caption{A cluster of LEO satellites traverses orbit while simultaneously delivering positioning and communication services to UTs. Downlink positioning occurs in the large-timescale, whereas JUDE is carried out in the small-timescale.}
    \label{fig:system}
\end{figure}

As illustrated  in Fig. \ref{fig:system}, we consider a cluster of $S$ LEO satellites simultaneously providing downlink positioning and uplink communication services to $U$ UTs. Each satellite is equipped with a UPA having $M=M_h \times M_v$ antennas, where $M_h$ and $M_v$ denote the numbers of antennas in the horizontal and vertical dimensions, respectively. The spacing between the antennas is set to one-half wavelength. Each UT is equipped with a single antenna. 
\subsection{Channel Models} \label{ch_model}
Since both downlink positioning and uplink communication are considered, we first present the time-frequency channel model used for downlink positioning, followed by a discussion of the tap delay line channel model employed for uplink channel estimation.
\subsubsection{Time-frequency Channel}
We consider a downlink orthogonal frequency division multiplexing (OFDM) SATCOM system with $K$ subcarriers. Let $\Delta f =B/K$ and $T = 1/\Delta f$ denote the subcarrier spacing and OFDM symbol duration, respectively, where $B$ is the signal bandwidth. According to the Rician channel mode adopted in \cite{12,13,14,15}, the channel $\mathbf{h}_{s,u}(t,k)$ from the $s$-th LEO satellite to the $u$-th UT during the $t$-th OFDM symbol over the $k$-th subcarrier, can be decomposed into LoS and non line-of-sight (NLoS) components as follows
\begin{equation}
\begin{aligned}
    \mathbf{h}_{s,u}(t,k) &= \mathbf{h}_{s,u}^{\text{LoS}}(t,k) + \mathbf{h}_{s,u}^{\text{NLoS}}(t,k), \\
    \mathbf{h}_{s,u}^{\text{LoS}}(t,k) &= \sqrt{\frac{\beta_{s,u} \kappa_u}{\kappa_u + 1}} e^{j2\pi(tv_{s,u,0} - k\Delta f\tau_{s,u,0})} \mathbf{a}(\boldsymbol{\theta}_{s,u,0}), \\
    \mathbf{h}_{s,u}^{\text{NLoS}}(t,k) &= \sqrt{\frac{\beta_{s,u}}{\kappa_u + 1}} \sqrt{\frac{1}{P-1}} \sum_{p \ne 0} g_{s,u,p}\\&\times e^{j2\pi(tv_{s,u,p} - k\Delta f\tau_{s,u,p})}\mathbf{a}(\boldsymbol{\theta}_{s,u,p}),
\end{aligned}
\end{equation}
where $P$ denotes the number of propagation paths with $p=0$ representing the LoS path, $g_{s,u,p}$ is the complex path gain, $v_{s,u,p}$ the Doppler shift, and $\tau_{s,u,p}$ the propagation delay. $\beta_{s,u}$ and $\kappa_u$ are the large-scale fading coefficient and Rician $K$-factor, respectively. $\mathbf{a}(\boldsymbol{\theta}_{s,u,p}) \in \mathbb{C}^{M\times 1}$ is the array response vector at the satellite, where $\boldsymbol{\theta}_{s,u,p}=[\theta_{s,u,p} ^{az}, \theta_{s,u,p} ^{el}]^T$ denotes the angle-of-departure (AoD), comprising both azimuth and elevation angles. Let $\mathbf{m}(M)=[0,\cdots,M-1]^T$. The array response vector is given by
\begin{equation}
    \mathbf{a}(\boldsymbol{\theta}_{s,u,p}) = e^{-j2\pi \phi_{s,u,p}^{h}\mathbf{m}(M_h)} \otimes e^{-j2\pi \phi_{s,u,p}^{v}\mathbf{m}(M_v)},
\end{equation}
where $\phi_{s,u,p}^{h}=d\cos\theta_{s,u,p}^{az}\cos\theta_{s,u,p}^{el} /\lambda$ and $\phi_{s,u,p}^{v}=d\sin\theta_{s,u,p}^{az}\cos\theta_{s,u,p}^{el} /\lambda$. Here, $d$ is the antenna spacing along each dimension, and $\lambda$ denotes the wavelength corresponding to the central carrier frequency.

The large-scale fading coefficient $\beta_{s, u}$, in the dB domain, is given as \cite{16,17,18}
\begin{equation}
\begin{aligned}
-10 \log _{10} \beta_{s, u}=\beta_{s, u}^{\mathrm{FS}}+\beta_{s, u}^{\mathrm{SF}}+\beta_{s, u}^{\mathrm{CL}}+\beta_{s, u}^{\mathrm{AB}}+\beta_{s, u}^{\mathrm{SC}} \, [\mathrm{dB}].
\end{aligned}
\label{beta}
\end{equation}
Here, $\beta_{s, u}^{\mathrm{FS}}$ is the free-space path loss, $\beta_{s, u}^{\mathrm{SF}}$ is the shadow fading loss represented by a Gaussian random variable, $\beta_{s, u}^{\mathrm{CL}}$ represents the clutter loss, $\beta_{s, u}^{\mathrm{AB}}$ captures atmospheric absorption effects, and $\beta_{s, u}^{\mathrm{SC}}$ is the attenuation due to ionospheric or tropospheric scintillation.
We now examine key characteristics of the LEO satellite channel and highlight their differences from those of conventional terrestrial channels \cite{11}.
\begin{itemize}
\item \textit{Delay:} Due to the large distance between satellite and UTs, the propagation delay $\tau_{s,u,p}$ is large, but the delay spread $\Delta\tau_{s,u,p}$ is small due to limited scattering.

\item\textit{Angle:} Unlike terrestrial channels with diverse AoDs, all paths in LEO channel share the same AoD as the LoS path due to high orbit altitude, making the array response vectors identical across paths. Hence, $\mathbf{a}(\boldsymbol{\theta}_{s,u,p})=\mathbf{a}(\boldsymbol{\theta}_{s,u})$.

\item\textit{Doppler Shift:} Doppler shifts in LEO satellite channels $v_{s,u,p}$ originate from both satellite and UT mobility: $v_{s,u,p} = v^{\text{sat}}_{s,u,p} + v^{\text{ut}}_{s,u,p}$. The satellite-induced Doppler shift $v^{\text{sat}}_{s,u,p}$ is common across paths due to the satellite's high altitude, while UT-induced Doppler shift $v^{\text{ut}}_{s,u,p}$ varies with the UT's motion and local scattering. Modeling of $v^{\text{ut}}_{s,u,p}$ can be similar to that in terrestrial communications.
\end{itemize}
Keeping in view the above characteristics, the overall channel model becomes
\begin{equation}
    \mathbf{h}_{s,u}(t,k) = \sqrt{\frac{\beta_{s,u}}{\kappa_u + 1}}  G_{s,u}(t,k) \mathbf{a}(\boldsymbol{\theta}_{s,u}),
    \label{eq:channel}
\end{equation}
where
\begin{equation*}
\begin{aligned}
    G_{s,u}(t,k) &= \sqrt{\kappa_u} e^{j2\pi(tv_{s,u,0} - k\Delta f\tau_{s,u,0})} \\ 
    &+ \sqrt{\frac{1}{P-1}} \sum_{p \ne 0} g_{s,u,p} e^{j2\pi(tv_{s,u,p} - k\Delta f\tau_{s,u,p})}.
\end{aligned}
\end{equation*}

\subsubsection{Tap Delay Line Channel}
For uplink channel estimation, the tap delay line channel model is employed. Moreover, since the velocity of LEO satellites can be obtained from the orbital ephemerides contained within the two-line element (TLE) files \cite{celestrak2024tle,kassas2024ad}, the satellite-induced Doppler shift can be compensated \cite{liu2022robust}. Similarly, by using the known position of the satellite and the UT’s position information obtained during the downlink phase, the LoS delay can be compensated. The equivalent tap delay line channel model corresponding to (\ref{eq:channel}), after Doppler and delay compensation, can be written as \cite{li2023channel,bjornson2024introduction}
\begin{equation}
\mathbf{h}_{s,u}(t, \tau) = \sum_{p=0}^{P-1} \boldsymbol{\alpha}_{s,u,p}(t) \, \delta(\tau - \Delta\tau_{s,u,p}),
\label{tapdelaymodel}
\end{equation}
where $\Delta\tau_{s,u,p}=\tau_{s,u,p}-\tau_{s,u,0}$, $\boldsymbol{\alpha}_{s,u,p}(t)=g_{s,u,p}(t)\mathbf{a}(\boldsymbol{\theta}_{s,u}) \in\mathbb{C}^{M\times1}$, and $g_{s,u,p}(t)$ are the time-variant channel taps. As illustrated in Fig. \ref{fig:frame}, time-division duplexing (TDD) is adopted to differentiate between uplink and downlink transmissions. Consequently, by exploiting uplink-downlink channel reciprocity, the AoD and angle-of-arrival (AoA) remain identical in (\ref{eq:channel}) and (\ref{tapdelaymodel}). It is important to highlight that the proposed approach doesn't necessitate the use of TDD and can also operate under frequency-division duplexing (FDD), since it exploits only geometric reciprocity, which holds in FDD systems as well. The channel model of (\ref{tapdelaymodel}) is adopted for uplink channel estimation since we need a state-space formulation for the implementation of KF-based channel tracking. Luckily, the state evolution of the channel taps $g_{s,u,p}(t)$ can be modeled by a first-order autoregressive (AR) process by approximating the nonrational Jakes model \cite{jakes,bello1963characterization}. Detailed discussion of the state-space model will be presented in Section \ref{ssmodel}. After getting the estimates of the channel taps $\hat{g}_{p}(t)$ from the KF, the uplink time-frequency channel can be obtained by taking the discrete Fourier transform (DFT) of the tap delay line channel model as follows \cite{li2023channel,bjornson2024introduction}
\begin{equation}
\mathbf{h}_{s,u} (t,k)= \sum_{p=0}^{P-1} \hat{g}_{s,u,p}(t) e^{-j2\pi k\Delta f \Delta\tau_{s,u,p}} \mathbf{a}(\boldsymbol{\theta}_{s,u}).
\label{reconst}
\end{equation}
Since the array response vectors can be directly determined from the UTs' position information, the full channel vectors can be reconstructed once the scalar channel gains are accurately estimated.

\subsection{Frame Structure} \label{frame}
The LEO satellite channel of (\ref{eq:channel}) exhibits a natural separation into large- and small-timescale parameters \cite{zhang2025positioning}. Large-timescale parameters are associated with the positions of the UTs and satellites. Since the position of UTs vary slowly and satellite orbits are predictable and well-determined, large-timescale parameters can be accurately predicted. They include Doppler shift, delay, and AoD. In contrast to geometry-determined large-timescale parameters, small-timescale parameters, which consist of random variations in channel amplitude and phase, are inherently unpredictable and fluctuate rapidly. The motivation for using a dual-timescale approach is that it allows leveraging position information from the large-timescale while estimating the uplink channel on the small-timescale. Since the position remains coherent, only the scalar channel gain needs to be estimated, and the steering vectors can be reconstructed from the UTs' position information. Consequently, vector-type channel estimation is effectively reduced to scalar-type estimation, potentially improving channel estimation accuracy without increasing pilot overhead.

\begin{figure}
    \centering
    \includegraphics[width=0.5\textwidth]{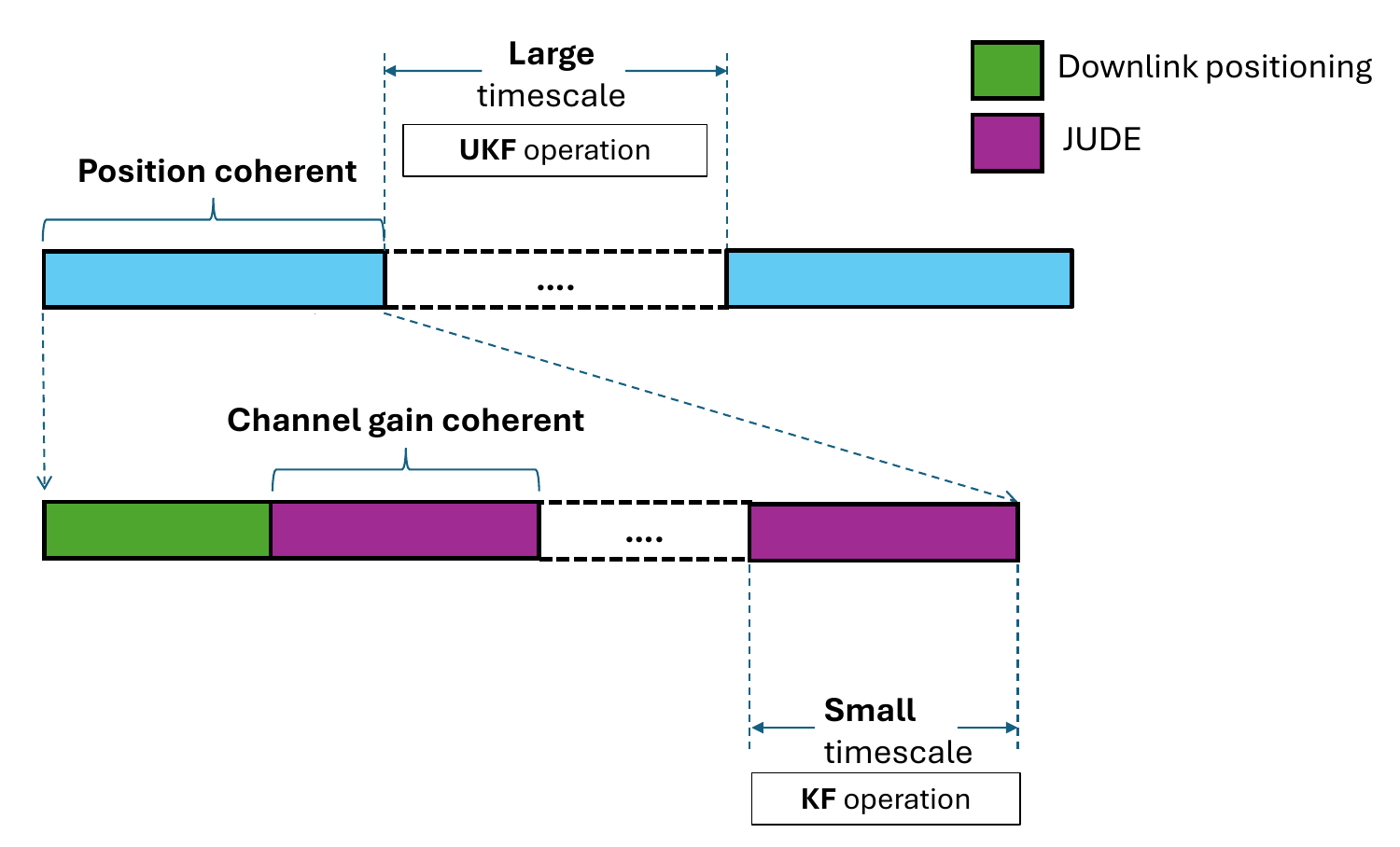}
    \caption{Frame structure and timescale of the proposed IPAC framework.}
    \label{fig:frame}
\end{figure}

Motivated by this analysis, a two-timescale frame structure is employed to integrate downlink positioning and uplink communication. The frame structure is depicted in Fig. \ref{fig:frame}. The proposed Kalman filters operate on two-timescale basis. Each large-timescale frame is governed by the position-coherent interval, which indicates the interval after which the UTs’ position information becomes outdated. It begins with a downlink positioning subframe where the UKF is employed to track UTs' position and velocity and whose estimates are then fed back to the LEO satellites. Multiple small-timescale subframes follow during which the channel gain remains coherent. The second KF is used during each small-timescale subframe for uplink channel estimation followed by cooperative data detection. TDD is employed to differentiate between uplink and downlink transmissions; however, as stated earlier, both TDD and FDD can be used. The proposed approach exploits uplink-downlink geometric reciprocity by employing UTs' positioning information from the downlink phase to improve joint channel estimation and data detection during the uplink phase.

The following two sections provide a detailed description of the proposed IPAC framework. First, the details of the downlink UT positioning part are presented, followed by a discussion on joint channel tracking and cooperative data detection during the uplink phase.

\section{PROPOSED IPAC FRAMEWORK: UKF FOR DOWNLINK POSITIONING}
During the downlink positioning phase, orthogonal frequency division is adopted for different satellites so that multiple satellites can transmit simultaneously. The signal received at the $u$-th UT from the $s$-th LEO satellite during the $t$-th OFDM symbol over the $k$-th subcarrier is given by
\begin{equation}
    y_{s,u}(t,k) = \mathbf{h}_{s,u}^H(t,k)\mathbf{F}_s\mathbf{s}(t,k) + n_{s,u}(t,k),
    \label{eq:rx signal}
\end{equation}
where $\mathbf{h}_{s,u}(t,k) \in \mathbb{C}^{M \times 1}$ is the channel vector given by (\ref{eq:channel}), $\mathbf{F}_s \in \mathbb{C}^{M\times U}$ is the analog precoding matrix, $\mathbf{s}(t,k)\in \mathbb{C}^{U\times 1}$ is the transmitted signal, and $n_{s,u}(t,k) \sim \mathcal{CN}(0, \sigma^2)$ is the circular symmetric complex additive white Gaussian noise (AWGN).

Due to the non-linearity of the measurement model, as described later in this section, the UKF is employed to track UT's position and velocity \cite{zheng2024leo}. Unlike the extended Kalman filter (EKF), the UKF avoids linearization and explicit derivative computation by using a set of sigma points to approximate the first- and second-order moments of the state distribution \cite{19,20}. The state-space model and the UKF implementation are discussed next.
\subsection{State-Space Model for UKF}
Let us denote the unknown 3D position and 3D velocity of a particular UT as $\mathbf{p} \in \mathbb{R}^3,\mathbf{v} \in \mathbb{R}^3$. To formulate the tracking problem, the unknown state vector is defined as
\begin{equation}
\boldsymbol{\zeta}=\left[\mathbf{p}^{\top}, \mathbf{v}^{\top}, \mathbf{b}^{\top}\right]^{\top} \in \mathbb{R}^{S+6}
\end{equation}
where $\mathbf{b}=\left[b_1, b_2, \ldots, b_S\right]^{\top} \in \mathbb{R}^S$ is the unknown clock bias between the LEO satellites and UT. The measurements, on the other hand, are the desired LoS channel parameters between each satellite and the UT. The combined measurement vector can be written as
\begin{equation}   \boldsymbol{\rho}=\left[\boldsymbol{\rho}_1^{\top}, \ldots, \boldsymbol{\rho}_S^{\top}\right]^{\top} \in \mathbb{R}^{2S},
    \label{eq:rho}
\end{equation}
where $\boldsymbol{\rho}_s=\left[v_{s,u,0}, \tau_{s,u,0} \right]^{\top} \in \mathbb{R}^2$ represents the measurement vector corresponding to $s$-th satellite, with $v_{s,u,0}$ and $\tau_{s,u,0}$ denoting the LoS Doppler shift and delay, respectively.

We now incorporate time indices into the state vector $\boldsymbol{\zeta}$ and the measurement vector $\boldsymbol{\rho}$.  Accordingly, the sequence of measurements over discrete time instances is represented as $\left\{\boldsymbol{\rho}_0, \boldsymbol{\rho}_1, \ldots, \boldsymbol{\rho}_N\right\}$, with the corresponding state sequence denoted by $\left\{\boldsymbol{\zeta}_0, \boldsymbol{\zeta}_1, \ldots, \boldsymbol{\zeta}_N\right\}$. This leads to the formulation of the following nonlinear dynamical system
\begin{equation}
\begin{aligned}
  \text { Time update:} \quad \boldsymbol{\zeta}_{n+1}=f\left(\boldsymbol{\zeta}_n\right),\\
    \text {Measurement update:} \quad \boldsymbol{\rho}_n=h\left(\boldsymbol{\zeta_n}\right). 
\end{aligned}
\end{equation}
where $f(\cdot)$ and $h(\cdot)$ are the process and measurement functions, respectively.

\subsubsection{Process Function} $f(\cdot)$ represents the evolution of state vector over time through the following relationships:
    \begin{equation}
    \begin{aligned}
    & \mathbf{p}_{n+1}=\mathbf{p}_n+\mathbf{v}_n \delta t+\hat{\mathbf{a}}_n \frac{\delta t^2}{2}, \\
    & \mathbf{v}_{n+1}=\mathbf{v}_n+\hat{\mathbf{a}}_n \delta t, \\
    & \mathbf{b}_{n+1}=\mathbf{b}_n.
    \end{aligned}
    \label{eq:process}
    \end{equation}    
Here, $\delta t$ represents the update interval and $\hat{\mathbf{a}}_n \in \mathbb{R}^3$ is the UT's acceleration at the $n$-th time instant. It is assumed that the acceleration measurement can be obtained from an inertial measurement unit (IMU) in the UT with some error $\tilde{\mathbf{a}}_n \in \mathbb{R}^3$, and is given as 
    \begin{equation}   \hat{\mathbf{a}}_n=\overline{\mathbf{a}}_n+\tilde{\mathbf{a}}_n , 
    \end{equation}
where $\overline{\mathbf{a}}_n$ is the true value of UT's acceleration. The underlying uncertainty in the time update step can be represented as 
    \begin{equation}    \boldsymbol{\zeta}_{n+1}=f(\boldsymbol{\zeta}_n;\hat{\mathbf{a}}_n)+\Delta \boldsymbol{\zeta}_{n+1}, \end{equation}
where $\Delta \boldsymbol{\zeta}_{n+1}$ is the uncertainty in the updated state induced by the IMU measurement error $\tilde{\mathbf{a}}_n$. We model this error as an additive Gaussian noise $\tilde{\mathbf{a}}_n=\hat{\mathbf{a}}_n-\overline{\mathbf{a}}_n \sim \mathcal{N}(\mathbf{0}, \mathbf{C}_a)$, where $\mathbf{C}_a$ is the error covariance matrix.
    
Based on (\ref{eq:process}), the term containing $\hat{\mathbf{a}}_n$ act as source of perturbation to each component of the state vector. Hence, the uncertainty in the state update can be represented as 
    \begin{equation}
    \begin{aligned}
    &\quad \quad \quad \quad \Delta \boldsymbol{\zeta}_{n+1} \sim \mathcal{N}(\mathbf{0}, \mathbf{P}),\\
    \text { where }\\
    &\mathbf{P}=\left[\begin{array}{cccc}
    \frac{\delta_t^4}{4} \mathbf{C}_a & \frac{\delta_t^3}{2} \mathbf{C}_a & \mathbf{0}_{3 \times S}  \\
    \frac{\delta_t^3}{2} \mathbf{C}_a & \delta_t^2 \mathbf{C}_a & \mathbf{0}_{3 \times S}  \\
    \mathbf{0}_{S \times 3} & \mathbf{0}_{S \times 3} & \mathbf{0}_{S \times S} 
    \end{array}\right] \in \mathbb{R}^{(S+6) \times(S+6)}\\
    \end{aligned}
    \label{eq:P}
    \end{equation}
represents the covariance matrix of the uncertainty in the updated state.

\subsubsection{Measurement Function} \label{FIM} $h(\cdot)$ represents how the measurements, which are the LoS channel parameters between each satellite and the UT, map to the unknown UT state. The relationships between the geometric channel measurements and UT's state are presented next.

\paragraph{Doppler shift} The Doppler shift observed from the LoS channel between the $s$-th LEO satellite and the $u$-th UT can be defined as
\begin{equation}
       v_{s,u,0}=\frac{\left(\mathbf{v}_s-\mathbf{v}\right)^{\top}\left(\mathbf{p}-\mathbf{p}_s\right)}{\lambda\left\|\mathbf{p}-\mathbf{p}_s\right\|_2}. 
\end{equation}
where $\mathbf{v}_s \in \mathbb{R}^3$ and $\mathbf{p}_s \in \mathbb{R}^3$ represent the velocity and position of the $s$-th satellite, respectively.
 \paragraph{Delay} The time delay of the LoS channel between the $s$-th satellite and the $u$-th UT can be written as
 \begin{equation}
        \tau_{s,u,0}=\frac{\left\|\mathbf{p}_s-\mathbf{p}\right\|_2}{c}+b_s.
 \end{equation}    
where $c$ denotes the speed of light and $b_s$ is an unknown but fixed clock bias between the $s$-th satellite and the UT.

The underlying uncertainty in the measurement update step can be represented as 
    \begin{equation}    \boldsymbol{\rho}_{n}=h(\boldsymbol{\zeta}_n)+\Delta \boldsymbol{\rho}_{n}, 
    \end{equation}
where $\Delta \boldsymbol{\rho}_{n}$ represents the estimation error of the LoS channel parameters $\boldsymbol{\rho}_{n}$. Assuming an efficient channel estimator, the covariance matrix of $\Delta \boldsymbol{\rho}_{n}$ is obtained from the inverse of the Fisher information matrix (FIM) of the corresponding channel estimation problem \cite{kay1993fundamentals}.

 \paragraph{Derivation of FIM} We can rewrite the received signal in (\ref{eq:rx signal}) as the sum of the LoS signal $\ell_{s,u}(t,k)$, all the NLoS components $z_{s,u}(t,k)$, and noise as follows
\begin{equation}
\begin{aligned}
    y_{s,u}(t,k) &= \sqrt{\frac{\beta_{s,u}\kappa_u}{\kappa_u + 1}} e^{-j2\pi(t v_{s,u,0} - k \Delta f \tau_{s,u,0})} \mathbf{a}^H(\boldsymbol{\theta}_{s,u}) f_s(t,k)\\
    &+\sqrt{\frac{\beta_{s,u}}{(\kappa_u + 1)(P - 1)}} \sum_{p \ne 0} g_{s,u,p}^* e^{-j2\pi(t v_{s,u,p} - k \Delta f \tau_{s,u,p})}\\ &\times\mathbf{a}^H(\boldsymbol{\theta}_{s,u}) f_s(t,k)\\
    &= \ell_{s,u}(t,k) + z_{s,u}(t,k)+ n_{s,u}(t,k).
\end{aligned}
\end{equation}
Here, $f_s(t,k)=\mathbf{F}_s\mathbf{s}(t,k)$ and $\mathbf{s}(t,k)$ is the unit modulus pilot signal. Since the NLoS components undergo Rayleigh fading, they can be modeled as a zero-mean circularly symmetric complex Gaussian random variable denoted by $\alpha_{s,u}$. Moreover, since $n_{s,u}(t,k) \sim \mathcal{CN}(0, \sigma^2)$, we can conclude that $z_{s,u}(t,k)+ n_{s,u}(t,k) \sim \mathcal{CN}(0, C_{s,u})$ where $C_{s,u}=\frac{\beta_{s,u}|\alpha_{s,u}|^2\|f_s(t,k)\|^2}{\kappa_u+1}+\sigma^2$.

The $\boldsymbol{\rho}$ defined in (\ref{eq:rho}) contains only the parameters of interest. For each satellite, we can collect the remaining unknown nuisance parameters as
$$ \boldsymbol{\eta}_s=\left[\boldsymbol{\theta}_{s,u}^{\top},\Re\{\alpha_{s,u}\}, \Im\{\alpha_{s,u}\})\right]^{\top} \in \mathbb{R}^{4}.
$$

Now, the LoS only version of received signals from the $s$-th satellite over $G$ pilots and $K$ subcarriers are stacked as    
    $\boldsymbol{\ell}_s=\left[\ell_{s,u}(1,1), \ldots, \ell_{s,u}(1,K), \ldots, \ell_{s,u}(g,1), \ldots,\right.\\ \left.\ell_{s,u}(g,K), \ldots,
    \ell_{s,u}(G,1), \ldots, \ell_{s,u}(G,K)\right]^{\top} \in \mathbb{C}^{G K}.$   
With the assumption of independent noise across different transmissions, subcarriers, and satellites, the following covariance matrix of the received signals is defined   
    $$
    \mathbf{C}=\operatorname{blkdiag}\left(\mathbf{C}_1, \ldots, \mathbf{C}_s, \ldots, \mathbf{C}_S\right),
    $$   
    where $ \mathbf{C}_s=\operatorname{diag}([C_{s,u}(1,1),\ldots,C_{s,u}(1,K),\ldots,C_{s,u}(g,1)\\,\ldots,C_{s,u}(g,K)\ldots,C_{s,u}(G,1),\ldots,C_{s,u}(G,K)]^{\top} ).$
    
Now, using the Slepian Bangs formula \cite{kay1993fundamentals}, the FIM of the channel parameters can be calculated as    
    \begin{equation}
        \mathbf{J}=2 \Re\left(\mathbf{D}^{H} \mathbf{C}^{-1} \mathbf{D}\right),
        \label{jacobian}
    \end{equation}      
where $\mathbf{D}\in \mathbb{R}^{GK\times (2S+4S)}$ is the Jacobian matrix computed as   
    $$
    \mathbf{D}=\left[\begin{array}{cccccc}
    \frac{\partial \boldsymbol{\ell}_1}{\partial \boldsymbol{\rho}_1} & \cdots & \mathbf{0} & \frac{\partial \boldsymbol{\ell}_1}{\partial \boldsymbol{\eta}_1} & \cdots & \mathbf{0} \\
    \vdots & \ddots & \vdots & \vdots & \ddots & \vdots \\
    \mathbf{0} & \cdots & \frac{\partial \boldsymbol{\ell}_S}{\partial \boldsymbol{\rho}_S} & \mathbf{0} & \cdots & \frac{\partial \boldsymbol{\ell}_S}{\partial \boldsymbol{\eta}_S}
    \end{array}\right] .
    $$
    
The contribution of nuisance parameters vectors $\boldsymbol{\eta}_s, s=1, \ldots, S$ to FIM $\mathbf{J}$ needs to be removed. Thus, we partition $\mathbf{J}$ as
$$
    \mathbf{J}=\left[\begin{array}{cc}
    \mathbf{X} & \mathbf{Y} \\
    \mathbf{Y}^{\top} & \mathbf{Z}
    \end{array}\right],
$$
where $\mathbf{X}\in\mathbb{R}^{2S\times2S}$ and $\mathbf{Z}\in\mathbb{R}^{4S\times4S}$. We can now compute the FIM of the desired channel domain parameters as   
    $$
    \mathbf{J}(\boldsymbol{\rho})=\mathbf{X}-\mathbf{Y} \mathbf{Z}^{-1} \mathbf{Y}^{\top} .
    $$
Finally, we can write
    \begin{equation}
        \Delta\boldsymbol{\rho}_n \sim \mathcal{N}(\mathbf{0},\boldsymbol{\Sigma}_n),
        \label{eq:Sigma}
    \end{equation}   
where $\boldsymbol{\Sigma}_n=\mathbf{J}^{-1}(\boldsymbol{\rho})$. 

\subsection{UKF Implementation}
The UKF is a variant of KF that eliminates the need for linearization and derivative computations. Instead, it approximates the first and second moments of the state estimates by generating a set of sigma points and propagating them through the non-linear state and measurement transformations to compute the sample mean and covariance \cite{20}. Once the state-space model is defined along with uncertainty representations, the time and measurement update steps of the proposed UKF are implemented as follows \cite{zheng2024leo}.

\subsubsection{Time Update} This step is used to predict the next state and update the covariance matrix by propagating the sigma points through the process function $f(.)$. Let $\boldsymbol{\bar{\zeta}}_{n-1}$ denote the mean of the state $\boldsymbol{\zeta}_{n-1}$ at time $n-1$, and $\mathbf{P}_{n-1}$ denote the corresponding covariance matrix of the perturbation. The first step is to generate $2L+1$ sigma vectors, where $L=S+6$ (dimensions of state vector), as follows
\begin{equation}
\begin{aligned}
\mathbf{s}_0 & =\boldsymbol{\bar{\zeta}}_{n-1}, \\
\mathbf{s}_\ell & =\boldsymbol{\bar{\zeta}}_{n-1} +\left(\sqrt{\left(L+\Lambda\right) \mathbf{P}_{n-1}}\right)_\ell, \quad \ell=1, \ldots, L, \\
\mathbf{s}_{\ell+L} & =\boldsymbol{\bar{\zeta}}_{n-1} -\left(\sqrt{\left(L+\Lambda\right) \mathbf{P}_{n-1}}\right)_\ell, \quad \ell=1, \ldots, L.
\end{aligned}
\end{equation}
Here, $(\sqrt{\mathbf{P}})_\ell$ denotes the $\ell$-th column of the square root of the matrix $\mathbf{P}$, and $\Lambda$ is a parameter controlling the dispersion of the generated sigma points.

These sigma vectors are then propagated through the process function to get $\tilde{\boldsymbol{\zeta}}_{n,\ell}=f(\mathbf{s}_\ell), \ell=0,1,\ldots,2L$. The mean and covariance matrix of the state estimate at time instant $n$ are predicted according to the following
\begin{equation}
\tilde{\boldsymbol{\zeta}}_n=\sum_{\ell=0}^{2 L} W_{\ell}^{(m)} \tilde{\boldsymbol{\zeta}}_{n, \ell} ,
\end{equation}
\begin{equation}
\tilde{\mathbf{P}}_n=\sum_{\ell=0}^{2 L} W_\ell^{(c)}\left(\tilde{\boldsymbol{\zeta}}_{n, \ell} - \tilde{\boldsymbol{\zeta}}_n\right)\left(\tilde{\boldsymbol{\zeta}}_{n, \ell} - \tilde{\boldsymbol{\zeta}}_n\right)^{\top}+\mathbf{P},
\end{equation}
where $\mathbf{P}$ is given by (\ref{eq:P}), and $W_\ell^{(m)}$ and $W_\ell^{(c)}$ are the associated weights for the corresponding sigma vector, which are typically set as 
$$
\begin{aligned}
W_0^{(m)} & =\frac{\Lambda}{L+\Lambda},\,\,\,\,
W_0^{(c)}  =\frac{\Lambda}{L+\Lambda}+\left(1-\alpha^2+\beta\right), \\
W_\ell^{(m)} & =W_\ell^{(c)}=\frac{1}{2\left(L+\Lambda\right)},\quad \ell=1,2, \ldots, 2 L, \\
\Lambda & =\alpha^2\left(L+\kappa\right)-L.
\end{aligned}
$$

\subsubsection{Measurement Update} This step is used to improve the predicted state vector by utilizing the latest measurement. To approximate the first and second moments of the measurement vector at time $n$, the predicted state mean $\tilde{\boldsymbol{\zeta}}_n$ and process covariance matrix $\tilde{\mathbf{P}}_n$ from the time update step are used to generate $2L+1$ sigma vectors 
\begin{equation}
\begin{aligned}
\mathbf{x}_0 & =\tilde{\boldsymbol{\zeta}}_n, \,\,\,\,\mathbf{x}_\ell =\tilde{\boldsymbol{\zeta}}_n +\left(\sqrt{\left(L+\Lambda\right) \tilde{\mathbf{P}}_n}\right)_\ell, \\
\mathbf{x}_{\ell+L} & =\tilde{\boldsymbol{\zeta}}_n -\left(\sqrt{\left(L+\Lambda\right) \tilde{\mathbf{P}}_n}\right)_\ell, \quad \ell=1, \ldots, L.
\end{aligned}
\end{equation}

These sigma vectors are then propagated through the measurement function to get $\tilde{\boldsymbol{\rho}}_{n,\ell}=f(\mathbf{x}_\ell), \ell=0,1,\ldots,2L$. The mean and covariance matrix of the measurement estimate at time instant $n$ are approximated according to the following
\begin{equation}
\tilde{\boldsymbol{\rho}}_n=\sum_{\ell=0}^{2 L} W_{\ell}^{(m)} \tilde{\boldsymbol{\rho}}_{n, \ell},
\end{equation}
\begin{equation}
\tilde{\mathbf{Q}}_n=\sum_{\ell=0}^{2 L} W_\ell^{(c)}\left(\tilde{\boldsymbol{\rho}}_{n, \ell} - \tilde{\boldsymbol{\rho}}_n\right)\left(\tilde{\boldsymbol{\rho}}_{n, \ell} - \tilde{\boldsymbol{\rho}}_n\right)^{\top}+\mathbf{\Sigma}_n,
\end{equation}
where $\mathbf{\Sigma}_n$ is given by (\ref{eq:Sigma}). As shown in section \ref{FIM}, $\mathbf{\Sigma}_n$ is the inverse of the FIM, which in turn, depends on the true UT state. However, the true state $\mathbf{\zeta}_n$ is typically not available and we can approximate $\mathbf{\Sigma}_n$ as \cite{zheng2024leo}
$$
\hat{\boldsymbol{\Sigma}}_n=\epsilon \mathbf{I}+\sum_{\ell=0}^{2 L} W_\ell^{(c)} \mathbf{J}^{-1}\left(\mathbf{x}_\ell\right),
$$
where $\epsilon$ is a regularization parameter which is inversely proportional to the level of trust in the calculated FIMs $\mathbf{J}\left(\mathbf{x}_i\right)$. Depending on the system configuration, when the dimension of $\mathbf{J}\left(\mathbf{x}_i\right)$ is higher, a lower value of $\epsilon$ yields better performance.
Aferwards, the Kalman gain is computed as
$$
\mathbf{K}_n=\tilde{\mathbf{Q}}_{\boldsymbol{\zeta \rho}, n} \tilde{\mathbf{Q}}_n^{-1},
$$
where $\tilde{\mathbf{Q}}_{\zeta \rho, n}$ denotes the cross-covariance matrix between the predicted state and observation and is given by
$$
\tilde{\mathbf{Q}}_{\boldsymbol{\zeta \rho}, n}=\sum_{\ell=0}^{2 L} W_\ell^{(c)}\left(\tilde{\boldsymbol{\zeta}}_{n, \ell} - \tilde{\boldsymbol{\zeta}}_n\right)\left(\tilde{\boldsymbol{\rho}}_{n, \ell}-\tilde{\boldsymbol{\rho}}_n\right)^{\top} .
$$

As a last step, the acquired observation vector $\boldsymbol{\rho}_n$ is utilized to improve the state estimation and covariance matrix as follows
\begin{equation}
\begin{aligned}
\hat{\boldsymbol{\zeta}}_{n} & =\tilde{\boldsymbol{\zeta}}_{n} + \mathbf{K}_n\left(\boldsymbol{\rho}_n-\tilde{\boldsymbol{\rho}}_n\right), \\
\hat{\mathbf{P}}_n & =\tilde{\mathbf{P}}_n-\mathbf{K}_n \tilde{\mathbf{Q}}_n \mathbf{K}_n^{\top}.
\end{aligned}
\end{equation}

\section{PROPOSED IPAC FRAMEWORK: JOINT UPLINK CHANNEL ESTIMATION AND COOPERATIVE DATA DETECTION (JUDE)}\label{JCEDD}
After successfully tracking the UTs' position and velocity using the UKF on the large-timescale, the obtained information is fed back to the LEO satellites, and leveraged to enhance JUDE on the small-timescale. Specifically, a KF-based JUDE approach is employed, where the KF not only enables accurate CSI acquisition but also reduces pilot overhead by exploiting information from data symbols. The two tasks of channel estimation and cooperative data detection are jointly addressed through the EM algorithm \cite{20}, in which data detection is carried out in the expectation step, while channel estimation is performed in the maximization step using the KF.

In the following, we present the signal model for multi-LEO satellite multiuser cooperative uplink communication, followed by the details of the KF-based channel estimation and the EM algorithm used to realize JUDE.
\subsection{Signal Model for Multi-LEO Satellite Cooperative Uplink Communication} \label{Up_rx_model}
Since the UT’s position is known from the downlink positioning phase and then fed back to the LEO satellites, the steering vector $\mathbf{a}(\boldsymbol{\theta}_{s,u})$ can be reconstructed. By exploiting the steering vector, maximum ratio combining (MRC) is performed at each satellite by defining the combiner as the hermitian of the steering vector .i.e. $\boldsymbol{w}_{s, u}=\mathbf{a}\left(\boldsymbol{\theta}_{s, u}\right)$. Moreover, the data detection in the uplink phase is accomplished cooperatively by combining the received signals at all LEO satellites from the $u$-th UT, at a central node. This cooperative detection leads to improved detection performance by increasing the received signal strength from a particular UT.
Let $s_{u}(t, k) \sim \mathcal{C} \mathcal{N}(0,1)$ be the symbol transmitted by the $u$-th UT, then the effective uplink channel after MRC at each satellite, between the $u$-th UT and $S$ satellites is given by 
\begin{equation}
\mathbf{h}_{\text {eff,} u}(t, k)=\left[
\boldsymbol{w}_{1, u}^H \mathbf{h}_{1, u}(t, k),\cdots,\boldsymbol{w}_{S, u}^H \mathbf{h}_{S, u}(t, k)\right]^T.
\label{eff_uplink_ch}
\end{equation}

Similarly, the effective interfering channel corresponding to all other UTs is given by
\begin{equation}
\mathbf{h}_{\text {eff,} u^{\prime}}(t, k)=\left[\boldsymbol{w}_{1, u}^H \mathbf{h}_{1, u^{\prime}}(t, k),\cdots,\boldsymbol{w}_{S, u}^H \mathbf{h}_{S, u^{\prime}}(t, k)
\right]^T.
\end{equation}

Then the combined signal, $\mathbf{y}(t, k)=\left[y_1(t, k),\cdots,y_S(t, k)\right]^T \in \mathbb{C}^{S \times 1}$,  received at $S$ LEO satellites from $U$ UTs at time instant $t$ over the $k$-th subcarrier becomes
\begin{equation}
\begin{aligned}
\mathbf{y}(t, k)&=\mathbf{h}_{\text {eff,} u}(t, k) s_u(t, k)+\sum_{u^{\prime} \neq u} \mathbf{h}_{\text {eff, } u^{\prime}}(t, k) s_{u^{\prime}}(t,k)\\&+\mathbf{n}(t, k),
\end{aligned}
\end{equation}
where $\mathbf{n}(t, k) \sim \mathcal{C} \mathcal{N}\left(\mathbf{0}, \sigma^2 \mathbf{I}_S\right)$. The inter-user interference (IUI) and noise $\mathbf{n}(t, k)$ are merged into the effective noise $\breve{\mathbf{n}}(t, k)$ as follows
\begin{equation}
\mathbf{y}(t, k)=\underbrace{\mathbf{h}_{\text {eff, } u}(t, k) s_u(t, k)}_ \text { Desired signal }+\underbrace{\breve{\mathbf{n}}(t, k)}_ \text { IUI+Noise }.
\label{IOeqn}
\end{equation}

Now, we will derive the statistics of the effective noise $\breve{\mathbf{n}}(t, k)$, which contains the IUI and actual noise. Since we know that
$s_{u^{\prime}}(t, k) \sim \mathcal{C} \mathcal{N}(0,1) \quad \text { and } \quad \mathbf{n}(t, k) \sim \mathcal{C} \mathcal{N}\left(\mathbf{0}, \sigma^2 \mathbf{I}_S\right)$,
we have $\breve{\mathbf{n}}(t, k) \sim \mathcal{C} \mathcal{N}\left(\mathbf{0}, \mathbf{R}_{\breve{\mathbf{n}}}\right)$, where
\begin{equation}
\begin{aligned}
\mathbf{R}_{\breve{\mathbf{n}}}&=E\left(\breve{\mathbf{n}}(t, k) \breve{\mathbf{n}}(t, k)^{H}\right) \\
&=\sum_{u^{\prime} \neq u} E\left(\mathbf{h}_{\text{eff,} u^{\prime}}(t, k) \mathbf{h}_{\text{eff,} u^{\prime}}(t, k)^{H}\right)+\sigma^2 \mathbf{I}_S \\
&=\operatorname{diag}(\sum_{u^{\prime} \neq u} \boldsymbol{w}_{s, u}^H \mathbf{a}\left(\boldsymbol{\theta}_{s, u^{\prime}}\right) E\left(h_{s, u^{\prime}}(t, k) h_{s, u^{\prime}}(t, k)^*\right)\\ &~\times\mathbf{a}^{H}\left(\boldsymbol{\theta}_{s, u^{\prime}}\right) \boldsymbol{w}_{s, u})+\sigma^2 \mathbf{I}_S \\
&=\operatorname{diag}(\sum_{u^{\prime} \neq u} \boldsymbol{w}_{s, u}^H \mathbf{a}\left(\boldsymbol{\theta}_{s, u^{\prime}}\right)(\sum_{p=0}^{P-1} \sigma_p^2) \mathbf{a}^{H}\left(\boldsymbol{\theta}_{s, u^{\prime}}\right) \boldsymbol{w}_{s, u})+\sigma^2 \mathbf{I}_S.
\end{aligned}
\end{equation}
Here, $h_{s, u^{\prime}}(t, k)=\sqrt{\frac{\beta_{s,u^\prime}}{\kappa_{u^\prime + 1}}}  G_{s,u^\prime}(t,k)$, and $\sigma_p^2=\frac{\beta_{s,{u^\prime}}\kappa_{u^\prime}}{\kappa_{u^\prime} + 1}$ is the variance of each channel tap evolving according to the first-order autoregressive model of (\ref{timefbkf}). The details of the KF employed for uplink channel estimation are presented in the following subsection.

\subsection{KF for Uplink Channel Estimation}
For uplink channel estimation using the KF, the tap delay line channel model of (\ref{tapdelaymodel}) is employed. Once the steering vector $\mathbf{a}(\boldsymbol{\theta}_{s,u})$ is obtained, instead of estimating the vector $\boldsymbol{\alpha}_{s,u,p}(t)$, only scalar channel tap gains $g_{s,u,p}(t)$ need to be estimated. Moreover, knowledge of the position and velocity of the UTs helps to find the LoS channel tap delay $\tau_{s,u,0}$ and Doppler frequency $v^{ut}_{s,u,0}$, which are essential for implementing the KF-based channel tracking approach. 

\subsubsection{State-Space Model} \label{ssmodel} Due to the mobile nature of the channel, the physical channel taps
$g_{s,u,p}(t)$ are time-variant. According to the wide-sense stationary and uncorrelated scattering (WSSUS) model \cite{bello1963characterization}, the channel taps are independent and each tap is a zero-mean wide-sense stationary complex Gaussian process with autocorrelation properties governed by the Doppler rate $v^{ut}_{s,u,0}T$. This model is based on approximating the nonrational Jakes model \cite{jakes} by a first-order AR model. We are interested in tracking the time variation of the channel taps corresponding to $u$-th UT and $S$ satellites, by using an AR(1) model as follows
\begin{equation}
\mathbf{g}_{t+1,u}= \mathbf{F} \mathbf{g}_{t,u}+\mathbf{G} \mathbf{u}_t
\label{timefbkf}
\end{equation}
where $\mathbf{g}_{t,u}=[g_{1,u,0}(t),\ldots,g_{s,u,p}(t),\ldots,g_{S,u,P-1}(t)]^T \in \mathbb{C}^{S P \times 1}$ and $\mathbf{u}_t \sim \mathcal{CN}(\mathbf{0},\mathbf{I}_{SP})$. $\mathbf{F} \in \mathbb{C}^{S P \times S P}$ and $\mathbf{G} \in \mathbb{C}^{S P \times S P}$ are block diagonal matrices defined as
\begin{equation*}
\begin{aligned}
\mathbf{F} &= \operatorname{blkdiag} \left(\alpha_1 \mathbf{I}_P , \alpha_2 \mathbf{I}_P,\cdots,\alpha_S \mathbf{I}_P \right), \\[1ex]
\mathbf{G} &= \operatorname{blkdiag}~(
\sqrt{(1-\alpha_1^2)\,C_1},\cdots,\sqrt{(1-\alpha_1^2)\,e^{-\eta (P-1)}C_1/\kappa_u}\\ 
& ,\cdots,\sqrt{(1-\alpha_S^2)\,C_S},\cdots,\sqrt{(1-\alpha_S^2)\,e^{-\eta (P-1)}C_S/\kappa_u}~),
\end{aligned}
\end{equation*}
where $\alpha_s=J_0\left(2 \pi v^{ut}_{s,u,0} T\right)$, $C_s=\frac{\beta_{s,u}\kappa_u}{\kappa_u + 1}$, and $J_0(.)$ is the zero-order Bessel function of the first kind. $\eta$ is the exponent of the channel decay profile and the factor $\sqrt{\left(1-\alpha_s^2\right)\,e^{-\eta p}C_s/\kappa_u}$ ensures that the exponential decay profile is maintained for all time. The covariance of $\mathbf{g}_{0,u}$ is given by $\mathbf{G}\mathbf{G}^H$.

Next, we will present the measurement update equation of the KF for the uplink multi-LEO satellite multiuser scenario. Let $\mathbf{Q}_{P}$ represents the first $P$ columns of $K\times K$ DFT matrix. First, the received signal at a single satellite and on a single subcarrier is presented. The signal received from $U$ UTs at the $s$-th satellite over the $k$-th subcarrier can be written as 
\begin{equation}
\mathbf{y}_s(t,k)= s_u(t,k)\mathbf{h}_{s,u}(t,k)+\sum_{u^{\prime} \neq u} s_{u^\prime}(t,k)\mathbf{h}_{s,u^\prime}(t,k)+\mathbf{n}_s(t,k).
\end{equation}
Plugging in the expression for $\mathbf{h}_{s,u}(t,k)$ from (\ref{reconst}), the signal becomes
\begin{equation}
\begin{aligned}
\mathbf{y}_s(t,k)&=s_{u}(t,k)(\mathbf{q}_k \otimes \mathbf{a}\left(\boldsymbol{\theta}_{s, u}\right))\mathbf{g}_{t, u}+\sum_{u^{\prime} \neq u} s_{u^\prime}(t,k)\\
&\times(\mathbf{q}_k \otimes \mathbf{a}\left(\boldsymbol{\theta}_{s, u^\prime}\right))\mathbf{g}_{t, u^\prime}+\mathbf{n}_s(t,k),
\end{aligned}
\end{equation}
where $\mathbf{q}_k$ is the $k$-th row of the DFT matrix $\mathbf{Q}_{P}$. Let $\boldsymbol{\mathcal { X }}^{u}=[s_u(t, 1),\cdots,s_u(t, K)]^T$ be the OFDM symbol transmitted from the $u$-th UT. The received signal over all $K$ subcarriers can be expressed as
\begin{equation}
\begin{aligned}
\mathbf{y}_s(t)&=([\operatorname{diag}\left(\boldsymbol{\mathcal { X }}^{u}\right) \mathbf{Q}_P] \otimes \mathbf{a}\left(\boldsymbol{\theta}_{s, u}\right))\mathbf{g}_{t, u}+\\&\sum_{u^{\prime} \neq u} ([\operatorname{diag}(\boldsymbol{\mathcal { X }}^{u^\prime}) \mathbf{Q}_P] \otimes \mathbf{a}\left(\boldsymbol{\theta}_{s, u^\prime}\right))\mathbf{g}_{t, u^\prime}+\mathbf{n}_s(t).
\end{aligned}
\end{equation}

Now, after performing MRC at $s$-th satellite the signal becomes
\begin{equation}
\begin{aligned}
\mathbf{y}_s(t)&=(\mathbf{I}_K \otimes \mathbf{a}^{H}\left(\boldsymbol{\theta}_{s, u}\right)([\operatorname{diag}\left(\boldsymbol{\mathcal { X }}^{u}\right) \mathbf{Q}_P] \otimes \mathbf{a}\left(\boldsymbol{\theta}_{s, u}\right))\mathbf{g}_{t, u}+\\&\sum_{u^{\prime} \neq u} (\mathbf{I}_K \otimes \mathbf{a}^{H}\left(\boldsymbol{\theta}_{s, u}\right)([\operatorname{diag}(\boldsymbol{\mathcal { X }}^{u^\prime}) \mathbf{Q}_P] \otimes \mathbf{a}\left(\boldsymbol{\theta}_{s, u^\prime}\right))\mathbf{g}_{t, u^\prime}\\&+\mathbf{n}_s(t).
\end{aligned}
\end{equation}

Lastly, the combined multi-LEO signal, $\mathbf{y}_t=\left[\mathbf{y}_1(t),\cdots,\mathbf{y}_S(t)\right]^T \in \mathbb{C}^{S K \times 1}$, received at the central node over all subcarriers $K$ at the time instant $t$ can be written as
\begin{equation}
\mathbf{y}_t=\mathbf{X}_{t, u} \mathbf{g}_{t, u}+\sum_{u^{\prime} \neq u} \mathbf{X}_{t, u^{\prime}} \mathbf{g}_{t, u^{\prime}}+\mathbf{n}_t,
\end{equation}
where $\mathbf{n}_{t}$ is the zero-mean complex Gaussian noise with covariance $\sigma_k^2 \mathbf{I}_{SK}$ and $\mathbf{X}_{t, u},\mathbf{X}_{t, u^{\prime}} \in \mathbb{C}^{S K \times S P}$
are defined as follows
\begin{equation}
\begin{aligned}
\mathbf{X}_{t, u}&=\operatorname{blkdiag}\left(\boldsymbol{X}_t^{\left(1, u\right)}, \cdots, \boldsymbol{X}_t^{\left(S, u\right)}\right), \\
\boldsymbol{X}_t^{\left(s, u\right)}&=\left(\mathbf{I}_K \otimes \mathbf{a}^{H}\left(\boldsymbol{\theta}_{s, u}\right)\right)\left(\left[\operatorname{diag}\left(\boldsymbol{\mathcal { X }}^{u}\right) \mathbf{Q}_P\right] \otimes \mathbf{a}\left(\boldsymbol{\theta}_{s, u}\right)\right),\\
\boldsymbol{X}_t^{\left(s, u^{\prime}\right)}&=\left(\mathbf{I}_K \otimes \mathbf{a}^{H}\left(\boldsymbol{\theta}_{s, u}\right)\right)([\operatorname{diag}(\boldsymbol{\mathcal { X }}^{u^{\prime}}) \mathbf{Q}_P] \otimes \mathbf{a}\left(\boldsymbol{\theta}_{s, u^{\prime}}\right)).
\end{aligned}
\label{Xt}
\end{equation}

The IUI and noise $\mathbf{n}_{t}$ are merged into the effective noise $\widetilde{\mathbf{n}}_t$ as follows
\begin{equation}
\mathbf{y}_t=\underbrace{\mathbf{X}_{t, u} \mathbf{g}_{t, u}}_\text{Desired signal}+\underbrace{\widetilde{\mathbf{n}}_t}_{\text{IUI + Noise}},
\label{measeqn}
\end{equation}
where $\widetilde{\mathbf{n}}_t=\sum_{u^{\prime} \neq u} \mathbf{X}_{t, u^{\prime}} \mathbf{g}_{t, u^{\prime}}+\mathbf{n}_t$.
Equations (\ref{timefbkf}) and (\ref{measeqn}) serve as the time and measurement update equations for the KF, respectively. Now, we will derive the statistics of effective noise $\widetilde{\mathbf{n}}_t$, which contains the IUI and actual noise $\mathbf{n}_{t}$. Based on whether we are sending known pilots or data, $\mathbf{X}_{t, u^{\prime}}$ can be deterministic or random. We will also make use of the prior knowledge about $\mathbf{g}_{t, u^{\prime}}$ and derive the statistics of effective noise $\widetilde{\mathbf{n}}_t$ for the case of deterministic $\mathbf{X}_{t, u^{\prime}}$ first.
Since $ \mathbf{n}_t \sim \mathcal{C N}\left(\mathbf{0}, \sigma^2 \mathbf{I}_{S K}\right)$ and from (\ref{timefbkf}), we can conclude that $\mathbf{g}_{t, u^{\prime}} \sim \mathcal{C} \mathcal{N}\left(\mathbf{c}_{t, u^{\prime}}, \mathbf{P}_{t, u^{\prime}}\right)$, hence $\widetilde{\mathbf{n}}_t \sim \mathcal{C} \mathcal{N}\left(E\left(\widetilde{\mathbf{n}}_t\right), \mathbf{R}_{\widetilde{\mathbf{n}}}\right)$. The statistics are computed as follows
\begin{equation}
\begin{aligned}
E\left(\widetilde{\mathbf{n}}_t\right)&=\sum_{u^{\prime} \neq u}E\left(\mathbf{X}_{t, u^{\prime}}\right) E\left(\mathbf{g}_{t, u^{\prime}}\right)+E\left(\mathbf{n}_t\right)\\
&=\sum_{u^{\prime} \neq u} \mathbf{X}_{t, u^{\prime}} \mathbf{c}_{t, u^{\prime}}, \text { (as } E\left(\mathbf{X}_{t, u^{\prime}}\right)=\mathbf{X}_{t, u^{\prime}})\\
\mathbf{R}_{\widetilde{\mathbf{n}}}&=E\left(\widetilde{\mathbf{n}}_t \widetilde{\mathbf{n}}_t^{H}\right)=\sum_{u^{\prime} \neq u} \mathbf{X}_{t, u^{\prime}} \mathbf{P}_{t, u^{\prime}} \mathbf{X}_{t, u^{\prime}}^{H}+\sigma^2 \mathbf{I}_{S K}.
\end{aligned}
\end{equation}

As $\mathbf{c}_{t, u^{\prime}}, \mathbf{P}_{t, u^{\prime}}$ are evolving with time so we don't know them in advance. Therefore, we will use the initial values $\mathbf{c}_{o, u^{\prime}}, \mathbf{P}_{o, u^{\prime}}$ instead. This assumption will obviously result in some perfomance loss but we are still able to get satisfactory results as presented in Section \ref{results}. The statistics for the case of random $\mathbf{X}_{t, u^{\prime}}$ are derived in Section \ref{max_step}.

\subsubsection{KF Implementation} After defining the state-space model, the KF can be implemented as follows.

For $i=1, \ldots, T$ , calculate
\begin{subequations}
\begin{align}
\mathbf{R}_{e, t}= & \mathbf{R}_{\widetilde{\mathbf{n}}}+ \mathbf{X}_t \mathbf{P}_{t \mid t-1}\mathbf{X}_t^H, \quad \mathbf{P}_{0 \mid-1}=\mathbf{G}\mathbf{G}^H, \label{kf1}\\
\mathbf{K}_t= & \mathbf{P}_{t \mid t-1} \mathbf{X}_t^H \mathbf{R}_{e, t}^{-1}, \label{kf2}\\
\hat{\mathbf{g}}_{t,u \mid t}= & \left(\mathbf{I}_{P}-\mathbf{K}_t \mathbf{X}_t\right) \hat{\mathbf{g}}_{t,u \mid t-1}+\mathbf{K}_t (\mathbf{y}_t -E(\widetilde{\mathbf{n}}_t)), \label{kf3}\\
\hat{\mathbf{g}}_{t+1,u \mid t}= & \mathbf{F} \hat{\mathbf{g}}_{t,u \mid t}, \quad \mathbf{g}_{0,u \mid-1}=[C_1,\mathbf{0}_{P-1}^T,\cdots,C_S,\mathbf{0}_{P-1}^T]^T, \label{kf4}\\
\mathbf{P}_{t+1 \mid t}= & \mathbf{F}\left(\mathbf{P}_{t \mid t-1}-\mathbf{K}_t \mathbf{R}_{e, t} \mathbf{K}_t^H\right) \mathbf{F}^H +\mathbf{G} \mathbf{I}_P \mathbf{G}^H. \label{kf5}
\end{align}
\label{fbkf1}
\end{subequations}
Here, the first equation (\ref{kf1}) computes the innovation covariance $\mathbf{R}_{e, t}$ from the measurement noise and predicted state uncertainty projected onto the observation space.
Equation (\ref{kf2}) computes the Kalman gain $\mathbf{K}_t$, which balances the predicted state and the new observation to update the estimate.
Equation (\ref{kf3}) performs the measurement update by correcting the predicted state with the innovation weighted by the Kalman gain.
Equation (\ref{kf4}) propagates the state estimate forward in time using the state transition matrix to perform the time update.
The last equation (\ref{kf5}) updates the error covariance for the next time step, including process noise and the effect of the Kalman gain correction.
The desired channel estimate is $\hat{\mathbf{g}}_{t,u \mid T}$. We can reconstruct the time-frequency channel from the estimated channel taps $\hat{\mathbf{g}}_{t,u \mid T}$ using (\ref{reconst}) and then use it for data detection in the expectation step of the EM algorithm discussed next.

\subsection{EM Algorithm for JUDE Implementation}
The EM algorithm is well-suited for JUDE because the received signal in (\ref{measeqn}) depends on two unknowns: the transmitted data symbols and the channel taps. Solving this joint problem becomes computationally intractable due to the large search space and the coupling between the data and the channel. EM provides an iterative approach that alternates between estimating the data (expectation step) given the current channel estimate, and refining the channel estimate (maximization step) using the detected data. This decomposition allows leveraging probabilistic structure, reduces complexity, and ensures that both tasks improve each other iteratively. Hence, to jointly accomplish the two tasks of channel estimation and cooperative data detection, the EM algorithm is employed \cite{fbkf,20}. The expectation step of the EM algorithm is used for data detection, while the maximization step is employed for channel estimation using the KF. The initial channel estimates used to initialize the EM algorithm are also obtained from the KF, utilizing pilot symbols. As we intend to do JUDE, we require two forms of the input/output equations for the EM algorithm: one that lends itself to channel estimation (i.e., treats the channel taps as the unknown) and a dual version that lends itself to data detection (i.e., treats the OFDM symbols as the unknown). The first input/output equation for channel estimation using the KF is (\ref{measeqn}).

\subsubsection{Expectation Step}The dual version of input/output equation for cooperative data detection at the central node is (\ref{IOeqn}). The data detection is done on a subcarrier basis. We will use (\ref{IOeqn}) in the expectation step to get the soft estimates (first and second moments) of the data symbols and consequently the soft estimates of $\mathbf{X}_{t,u}$. Given the current estimate of $\mathbf{g}_{t,u}$, the time-frequency channel is reconstructed using (\ref{reconst}) and the effective channel $\mathbf{h}_{\text{eff,} u}(t, k)$ in (\ref{IOeqn}) is obtained between $S$ satellites and $u$-th UT. Let $\mathcal{R}=\left\{r_1, \ldots, r_{|\mathcal{R}|}\right\}$ denote the alphabet set from which the transmitted symbols take their values. For example, for quadrature phase shift keying (QPSK) $\mathcal{R}$ has four possible symbols. We can evaluate the conditional probability distribution function (PDF) of the transmitted symbols $s_u(t,k) \in \mathcal{R}$ given the output $\mathbf{y}(t,k)$ and effective channel $\mathbf{h}_{\text{eff,} u}(t, k)$ using Bayes rule.
\begin{equation}
\begin{aligned}
& f\left(r_i \mid \mathbf{y}(t,k), \mathbf{h}_{\text{eff,} u}(t, k)\right)=\frac{f\left(r_i, \mathbf{y}(t,k) \mid \mathbf{h}_{\text{eff,} u}(t, k)\right)}{f\left(\mathbf{y}(t,k) \mid \mathbf{h}_{\text{eff,} u}(t, k)\right)} \\
&  \qquad =\frac{f\left(r_i, \mathbf{y}(t,k) \mid \mathbf{h}_{\text{eff,} u}(t, k)\right)}{\sum_{i=1}^{|\mathcal{R}|}f\left(\mathbf{y}(t,k),r_i \mid \mathbf{h}_{\text{eff,} u}(t, k)\right)} \\
&  \qquad=\frac{f\left(\mathbf{y}(t,k) \mid r_i,\mathbf{h}_{\text{eff,} u}(t, k)\right)f\left(r_i \mid \mathbf{h}_{\text{eff,} u}(t, k)\right)}{\sum_{i=1}^{|\mathcal{R}|} f\left(\mathbf{y}(t,k) \mid r_i,\mathbf{h}_{\text{eff,} u}(t, k)\right)f\left(r_i \mid \mathbf{h}_{\text{eff,} u}(t, k)\right)} \\
&  \qquad =\frac{e^{-\left((\mathbf{y}(t,k)-\mathbf{h}_{\text{eff,} u}(t, k)\,r_i)^H \mathbf{R}^{-1}_{\breve{\boldsymbol{n}}}(\mathbf{y}(t,k)-\mathbf{h}_{\text{eff,} u}(t, k)\,r_i)\right)}}{\sum_{i=1}^{|\mathcal{R}|} e^{e^{-\left((\mathbf{y}(t,k)-\mathbf{h}_{\text{eff,} u}(t, k)\,r_i)^H \mathbf{R}^{-1}_{\breve{\boldsymbol{n}}}(\mathbf{y}(t,k)-\mathbf{h}_{\text{eff,} u}(t, k)\,r_i)\right)}}} .
\end{aligned}
\end{equation}

We can use this conditional PDF to calculate the conditional expectation and second moment of $r_i$. 
\begin{equation}
\begin{aligned}
&E\left(r_i \mid \mathbf{y}(t,k), \mathbf{h}_{\text{eff,} u}(t, k)\right)= \\
&\frac{\sum_{i=1}^{|\mathcal{R}|} r_i e^{-\left((\mathbf{y}(t,k)-\mathbf{h}_{\text{eff,} u}(t, k)\,r_i)^H \mathbf{R}^{-1}_{\breve{\boldsymbol{n}}}(\mathbf{y}(t,k)-\mathbf{h}_{\text{eff,} u}(t, k)\,r_i)\right)}}{\sum_{i=1}^{|\mathcal{R}|} e^{-\left((\mathbf{y}(t,k)-\mathbf{h}_{\text{eff,} u}(t, k)\,r_i)^H \mathbf{R}^{-1}_{\breve{\boldsymbol{n}}}(\mathbf{y}(t,k)-\mathbf{h}_{\text{eff,} u}(t, k)\,r_i)\right)}}, \\
&E\left(|r_i|^2 \mid \mathbf{y}(t,k), \mathbf{h}_{\text{eff,} u}(t, k)\right)= \\
&\frac{\sum_{i=1}^{|\mathcal{R}|} r_i^2 e^{-\left((\mathbf{y}(t,k)-\mathbf{h}_{\text{eff,} u}(t, k)\,r_i)^H \mathbf{R}^{-1}_{\breve{\boldsymbol{n}}}(\mathbf{y}(t,k)-\mathbf{h}_{\text{eff,} u}(t, k)\,r_i)\right)}}{\sum_{i=1}^{|\mathcal{R}|} e^{-\left((\mathbf{y}(t,k)-\mathbf{h}_{\text{eff,} u}(t, k)\,r_i)^H \mathbf{R}^{-1}_{\breve{\boldsymbol{n}}}(\mathbf{y}(t,k)-\mathbf{h}_{\text{eff,} u}(t, k)\,r_i)\right)}} .
\end{aligned}
\label{moments_tone}
\end{equation}

Now, (\ref{moments_tone}) applies at a particular subcarrier frequency. So collecting (\ref{moments_tone}) for all tones $(k = 1, . . . , K)$ produces the two moments of the OFDM symbols $\boldsymbol{\mathcal { X }}^u$ in (\ref{Xt}). These moments are enough to characterize the first two moments of $\mathbf{X}_{t,u}$ required for channel estimation. Hence, we can derive that
\begin{equation}
 \begin{aligned}
  E(\mathbf{X}_{t,u}) =\operatorname{blkdiag}\left(\boldsymbol{\Phi}_1, \cdots, \boldsymbol{\Phi}_S\right),
\end{aligned}
 \label{soft2}
\end{equation}
where $\boldsymbol{\Phi}_s= \left(\operatorname{diag}\left[E(\boldsymbol{\mathcal { X }}^{u}\right] \mathbf{Q}_P\right) \otimes \left(\mathbf{a}^H\left(\boldsymbol{\theta}_{s, u})\mathbf{a}(\boldsymbol{\theta}_{s, u}\right)\right)$. The second moment is given by
\begin{equation}
 \begin{aligned}
  E(\mathbf{X}_{t,u} \mathbf{X}_{t,u} ^H) = \operatorname{blkdiag}\left[\mathbf{\Omega} _1,\mathbf{\Omega} _2,\ldots,\mathbf{\Omega} _S\right],
 \end{aligned}
 \label{soft3}
\end{equation}
where,
\begin{equation*}
 \begin{aligned}
\mathbf{\Omega} _s= (\mathbf{Q}_P^H \operatorname{diag}[E(\boldsymbol{\mathcal { X }}^u (\boldsymbol{\mathcal { X }}^u)^H)]\mathbf{Q}_P) \otimes(\mathbf{a}^H(\boldsymbol{\theta}_{s, u})\mathbf{a}(\boldsymbol{\theta}_{s, u})).
\end{aligned}
\end{equation*}

\subsubsection{Maximization Step} \label{max_step}
The maximazation step uses the KF again,  but this time with the soft estimates of $\mathbf{X}_{t,u}$ from the expectation step. Using the result from \cite{fbkf}, the channel estimate is now obtained using the KF equations (\ref{fbkf1}) but with the following substitutions
\begin{equation}
\begin{aligned}
 \mathbf{X}_{t,u} \rightarrow\left[\begin{array}{c}
E\left[\mathbf{X}_{t,u}\right] \\
\operatorname{Cov}\left[\mathbf{X}_{t,u}^H\right]^{1 / 2}
\end{array}\right], 
 \mathbf{y}_t \rightarrow\left[\begin{array}{c}
\mathbf{y}_t \\
\mathbf{0}_{SP}
\end{array}\right], 
\mathbf{I}_{SK} \rightarrow \mathbf{I}_{S(K+P)} .
\end{aligned}
\label{subs}
\end{equation}
This modification enforces the statistical uncertainty of symbol estimates in the channel update. The channel estimation version of the input/output equation (\ref{measeqn}) for the maximization step becomes
\begin{equation}
\begin{aligned}
\left[\begin{array}{c}
\mathbf{y}_t \\
\mathbf{0}_{SP}
\end{array}\right]&= \left[\begin{array}{c}
E\left[\mathbf{X}_{t,u}\right] \\
\operatorname{Cov}\left[\mathbf{X}_{t,u}^H\right]^{1 / 2}
\end{array}\right] \mathbf{g}_{t,u}+ \sum_{u^{\prime} \neq u}\left[\begin{array}{c}
E\left[\mathbf{X}_{t,u^{\prime}}\right] \\
\operatorname{Cov}\left[\mathbf{X}_{t,u^{\prime}}^H\right]^{1 / 2}
\end{array}\right]\\ &\times \mathbf{g}_{t,u^{\prime}}+\left[\begin{array}{c}\mathbf{n}_t \\ \mathbf{n}_v
\end{array}\right]
= \left[\begin{array}{c}
E\left[\mathbf{X}_{t,u}\right] \\
\operatorname{Cov}\left[\mathbf{X}_{t,u}^H\right]^{1 / 2}
\end{array}\right] \mathbf{g}_{t,u}+\widetilde{\mathbf{n}}_t,
\end{aligned}
\label{upd_meas}
\end{equation}
where $\widetilde{\mathbf{n}}_t$ is the effective noise containing both IUI and noise terms and $\mathbf{n}_v$ is virtual noise. Next, we will derive the statistics for the case of random $\mathbf{X}_{t, u^{\prime}}$ as here we don't know about the data symbols from other UTs contributing to the IUI term. Hence, we assume a uniform distribution based on the prior knowledge of the modulation scheme employed by UTs, that is we define $\mathbf{X}_{t, u^{\prime}} \sim U(\mathbf{0}, \mathrm{c}\mathbf{I}_{P\times P})$, where $\mathrm{c}=\mathbf{a}^H(\boldsymbol{\theta}_{s, u})\mathbf{a}(\boldsymbol{\theta}_{s, u^{\prime}})$. We know that $\mathbf{g}_{t, u^{\prime}} \sim \mathcal{C} \mathcal{N}\left(\mathbf{c}_{t, u^{\prime}}, \mathbf{P}_{t, u^{\prime}}\right)$, hence $\widetilde{\mathbf{n}}_t \sim \mathcal{C} \mathcal{N}\left(E\left(\widetilde{\mathbf{n}}_t\right), \mathbf{R}_{\widetilde{\mathbf{n}}}\right)$. The statistics are given as follows
\begin{equation}
\begin{aligned}
E\left(\widetilde{\mathbf{n}}_t\right)&=\sum_{u^{\prime} \neq u} E\left(\mathbf{X}_{t, u^{\prime}}^{a u g}\right) E\left(\mathbf{g}_{t, u^{\prime}}\right)+E\left(\mathbf{n}_t^{a u g}\right)\\
&=\sum_{u^{\prime} \neq u} \mathbf{X}_{t, u^{\prime}}^{a u g} \mathbf{c}_{t, u^{\prime}}, \quad\left(\text { as } E\left(\mathbf{X}_{t, u^{\prime}}^{a u g}\right)=\mathbf{X}_{t, u^{\prime}}^{a u g}\right)\\
\mathbf{R}_{\widetilde{\mathbf{n}}}&=E\left(\widetilde{\mathbf{n}}_t \widetilde{\mathbf{n}}_t^H\right)=\sum_{u^{\prime} \neq u} \mathbf{X}_{t, u^{\prime}}^{a u g} \mathbf{P}_{t, u^{\prime}} \mathbf{X}_{t, u^{\prime}}^{a u g H}+\sigma^2 \mathbf{I}_{S(K+P)},
\end{aligned}
\end{equation}
where $\mathbf{X}_{t, u^{\prime}}^{a u g}=\left[\begin{array}{c}
E\left[\mathbf{X}_{t,u^{\prime}}\right] \\
\operatorname{Cov}\left[\mathbf{X}_{t,u^{\prime}}^H\right]^{1 / 2}
\end{array}\right]$ and $\mathbf{n}_t^{a u g}=\left[\begin{array}{c}\mathbf{n}_t \\ \mathbf{n}_v
\end{array}\right]$.
As discussed earlier in Section \ref{ssmodel}, we will use the initial values of $\mathbf{c}_{t, u^{\prime}}, \mathbf{P}_{t, u^{\prime}}$.

\subsection{Summary of the Proposed JUDE Approach}
\begin{itemize}
    \item Calculate the initial channel estimate $\hat{\mathbf{g}}_{t,u}$ by using the KF (\ref{fbkf1}) with $\mathbf{X}_{t,u}$ and $\mathbf{y}_t$ corresponding to subcarriers with pilot symbols only.
    \item Iterate between expectation and maximization steps of EM algorithm for $n=1,2,\ldots,N_{iter}$
    \begin{itemize}
        \item Calculate the soft estimates of $\mathbf{X}_{t,u}$ using (\ref{moments_tone}), (\ref{soft2}), and (\ref{soft3}).
        \item Obtain the channel estimate $\hat{\mathbf{g}}_{t,u \mid T}$ by employing the KF with the substitutions of (\ref{subs}), i.e., using (\ref{upd_meas}) as the channel estimation version of the input/output equation.
    \end{itemize}
\end{itemize}
The algorithm terminates either upon reaching $N_{iter}$ iterations or when the change between successive estimates falls below a predefined threshold. Once EM iterations converge, the final soft estimates are demapped to the nearest constellation point and compared to ground-truth to compute the BER over all subcarriers and time slots.

\section{NUMERICAL RESULTS} \label{results}
\subsection{Simulation Setup}
\begin{figure}
    \centering
    \includegraphics[width=0.5\textwidth]{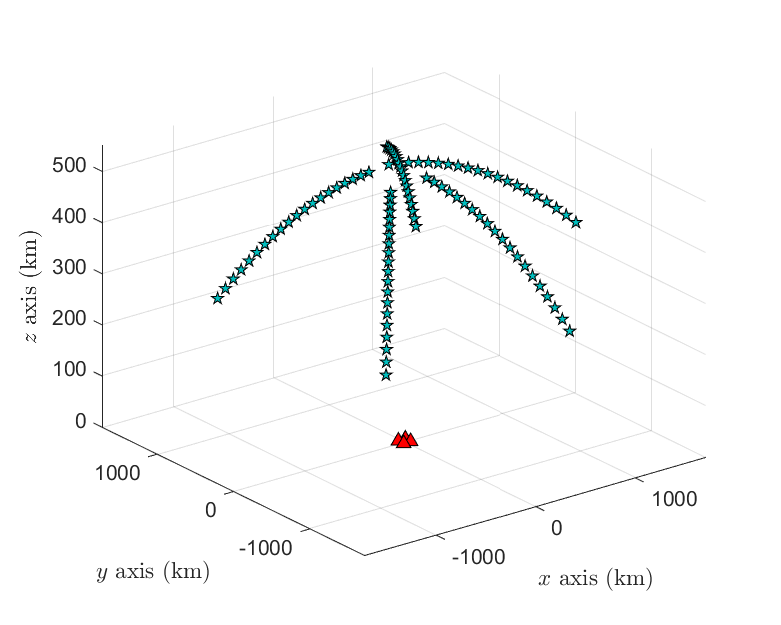}
    \caption{Constellation and trajectories of the 5 LEO satellites. Red triangles indicate positions of UTs.}
    \label{fig:const}
\end{figure}

During simulations, the Earth is modeled as a semi-sphere with radius 
$R_E=6400$ km, and a spherical coordinate system is defined with its origin at the Earth's center. The LEO satellites orbit at an altitude of 500 km. The constellation of LEO satellites and their trajectories relative to the area of interest --- where the UTs are located --- are generated using the QuaDRiGa toolbox \cite{quadriga}. The overall simulation setup is illustrated in Fig. \ref{fig:const}. The large-scale fading coefficient $\beta_{s, u}$ is calculated as per (\ref{beta}). The free space path loss and shadow fading loss are calculated as per the method provided in \cite{16}
\begin{equation}
\begin{aligned}
\beta_{s, u}^{\mathrm{FS}}&=20\log _{10}d+20\log _{10}f_c+32.45 [\mathrm{dB}],\\
\beta_{s, u}^{\mathrm{SF}}&=X(V_\sigma +V_\theta \log_{10}\theta_{sat})[\mathrm{dB}],
\end{aligned}
\end{equation}
where $f_c$ and $d=\left\|\mathbf{p}_u-\mathbf{p}_s\right\|_2$ are the carrier frequency (GHz) and distance (meters) between the satellite and the UT, respectively. $X \sim \mathcal{N}(0,1)$, $V_\sigma$ and $V_\theta$ are environment-specific parameters \cite{17}, and $\theta_{sat}$ is the satellite elevation angle in radians. $\beta_{s, u}^{\mathrm{AB}}$ depends on $f_c$ and $\theta_{sat}$, and is calculated according to the recommendations provided by the International Telecommunication Union in \cite{18,hourani2024atmospheric}. For LoS-dominant scenarios, $\beta_{s,u}^{\mathrm{CL}}=0~\mathrm{dB}$, while $\beta_{s,u}^{\mathrm{SC}}$ can be ignored for frequencies greater than 6 GHz \cite{16}. Other important simulation parameters are given in Table \ref{tab:sim_parameters}.

\begin{table}[!t]
\renewcommand{\arraystretch}{1.2}
\centering
\caption{\MakeUppercase{Simulation Parameters}}
\label{tab:sim_parameters}
\begin{tabular}{l c}
\toprule
\textbf{Parameter} & \textbf{Value} \\
\midrule
Number of LEO satellites $S$ & 5 \\
Number of UTs $U$ & 4 \\
Carrier frequency $f_c$ & 12.7 GHz (Ku band) \\
Subcarrier spacing $\Delta f$ & 120 kHz \\
Subcarrier number $K$ & 256 \\
Power budget at each LEO satellite & 50 dBm \\
Power budget at each UT & 30 dBm \\
Satellite array size $M = M_h \times M_v$ & $16 \times 16$ \\
Uplink modulation scheme & 16-PSK\\
\bottomrule
\end{tabular}
\end{table}

\subsection{Performance Evaluation of Proposed IPAC Framework}
Now, we will evaluate the performance of our approach by presenting and discussing the results for downlink positioning and JUDE, respectively.
\subsubsection{UKF for Downlink Positioning}
\begin{figure}
    \centering
    \includegraphics[width=0.5\textwidth]{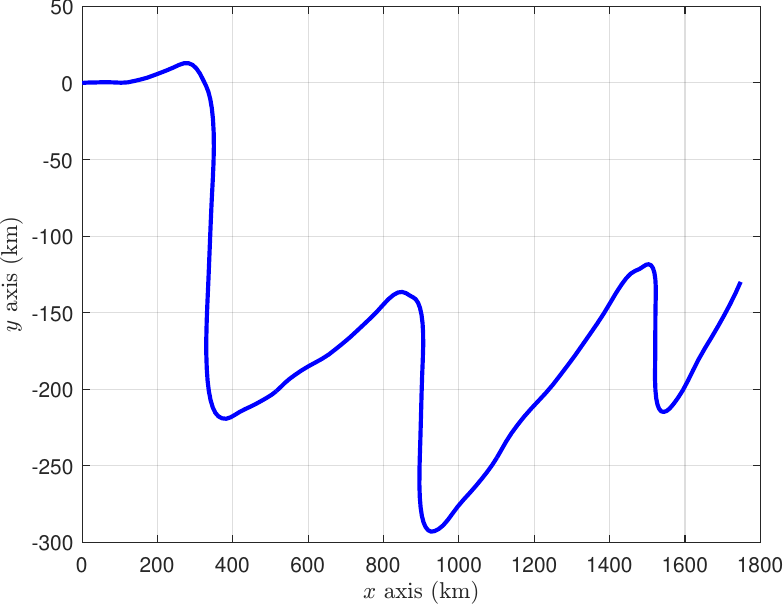}
    \caption{Simulated trajectory followed by one of the UTs during downlink positioning.}
    \label{fig:ut_path}
\end{figure}

\begin{figure}[!t]
    \centering
    \includegraphics[width=0.48\textwidth]{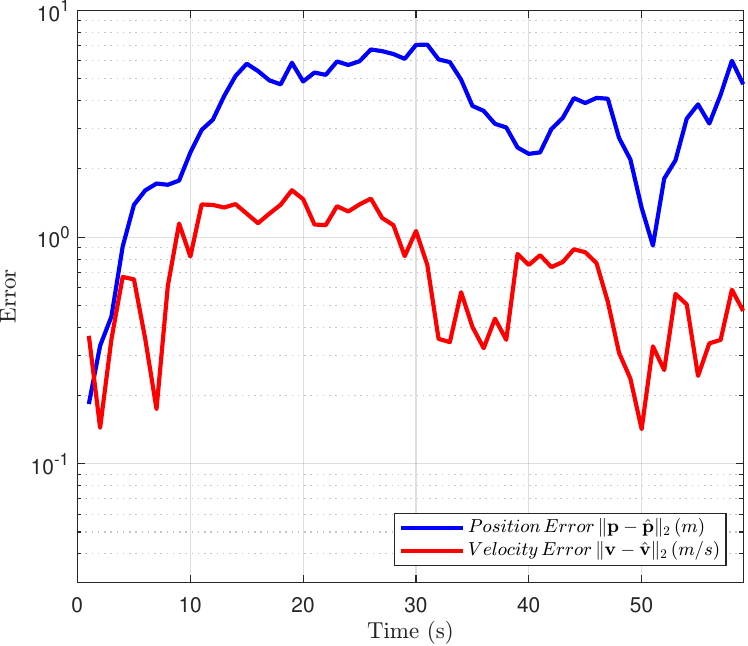}
    \caption{UKF tracking error for UT's position and velocity.}
    \label{fig:pos_vel_errors_comp}
\end{figure}

The results of downlink positioning accomplished using the UKF are depicted in Fig. \ref{fig:pos_vel_errors_comp}. We have used the tracking error in position and velocity as the metric for the evaluation of downlink positioning of UTs. Fig. \ref{fig:pos_vel_errors_comp} shows the errors in tracking position and velocity for a particular UT using the proposed UKF. The trajectory of the UT is also depicted in Fig. \ref{fig:ut_path}. It is evident from Fig. \ref{fig:pos_vel_errors_comp} that the proposed UKF effectively tracks the UT’s position and velocity with meter-level accuracy. The reason for slightly higher error in position as compared to velocity is that the UT's position is affected more from the error in acceleration measurement, which can be observed from (\ref{eq:process}).


\subsubsection{JUDE}
\begin{figure}[!t]
    \centering
    \includegraphics[width=0.49\textwidth]{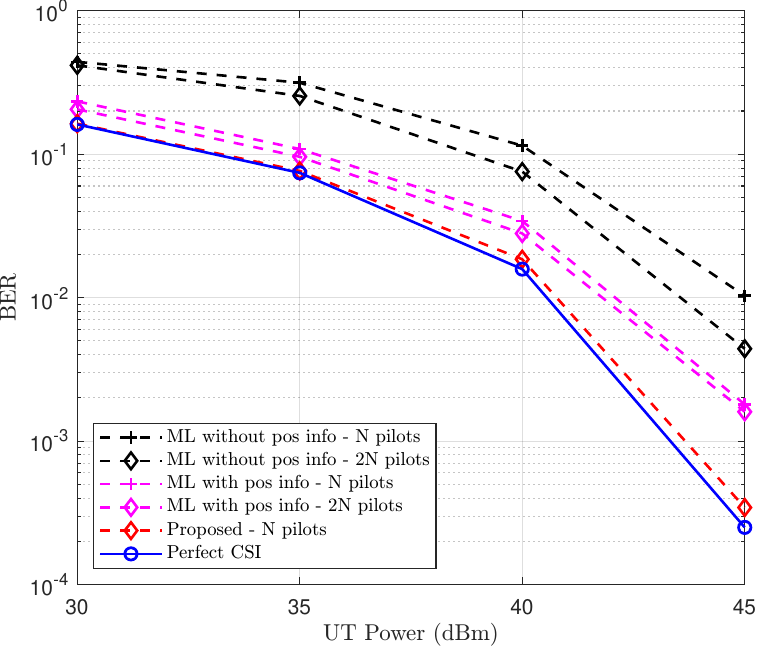}
    \caption{BER performance comparison of the proposed JUDE approach when the benchmark schemes are using the same number of pilots and twice the number of pilots.}
    \label{fig:BER1}
\end{figure}

After successful downlink positioning of UTs, the positioning information is fed back to the LEO satellites to aid in JUDE as explained in Section \ref{JCEDD}. We have employed two metrics for performance evaluation of proposed approach for JUDE. The bit error rate (BER) is used to evaluate data detection performance, whereas the channel normalized mean squared error (NMSE) is used to evaluate channel estimation performance and is defined as $\text{NMSE} = \frac{\mathbb{E}\!\left[\| \hat{\mathbf{h}} - \mathbf{h} \|^2\right]}{\mathbb{E}\!\left[\| \mathbf{h} \|^2\right]}
$.
\vspace{1em}
\paragraph{Benchmarking}
Figure \ref{fig:BER1} illustrates the uplink BER performance of the proposed scheme for the single UT case, benchmarked against the conventional pilot-based maximum likelihood (ML) approach and the perfect CSI case. Two ML-based schemes are considered for comparison: (i) an ML estimator that utilizes the available positioning information and estimates only the scalar channel gains rather than the full channel vector, ensuring a fair comparison with the proposed approach; and (ii) an ML estimator that operates without using positioning information, demonstrating the performance gain achieved by exploiting UTs' positioning information. It is evident from the Fig. \ref{fig:BER1} that ML approach without positioning information performs the worst highlighting the significance of IPAC for LEO satellite networks. The proposed IPAC approach not only outperforms the benchmark schemes but it also reduces the pilot overhead significantly leading to improved spectral efficiency. Fig. \ref{fig:BER1} depicts the superior BER performance of JUDE approach even when both the ML schemes are utilizing twice the number of pilots. 

\begin{figure}
    \centering
    \includegraphics[width=0.5\textwidth]{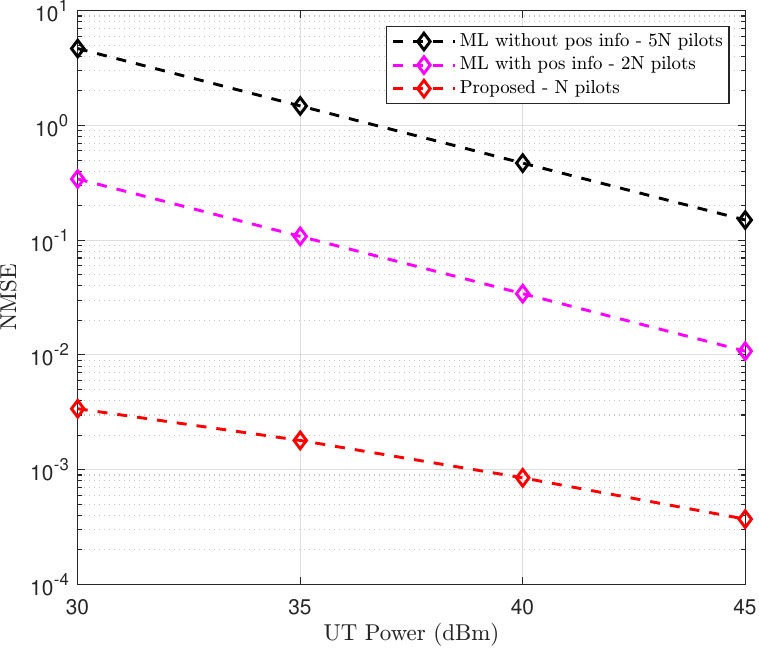}
    \caption{Channel NMSE performance comparison of the proposed JUDE approach with the benchmark schemes.}
    \label{fig:nmse}
\end{figure}

\begin{figure}
    \centering
    \includegraphics[width=0.5\textwidth]{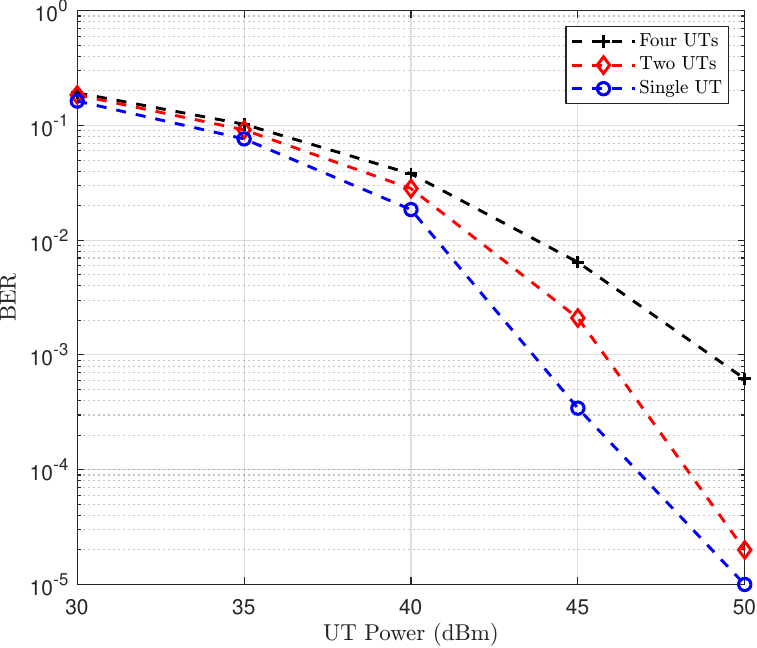}
    \caption{BER performance of the proposed JUDE approach for single and multiuser scenarios.}
    \label{fig:singlevsmulti}
\end{figure}

A similar performance gain is observed for the channel NMSE, as shown in Fig.~\ref{fig:nmse}. It is noteworthy that the benchmark schemes employ 2× and 5× more pilots than the proposed JUDE approach, yet still yield higher NMSE. This is because, in addition to leveraging positioning information, the proposed method performs channel estimation and data detection jointly, thereby exploiting the data symbols to further enhance channel estimation. Consequently, the ML scheme without positioning information requires significantly more pilots to match the performance of the JUDE approach.
\paragraph{Multiuser Scenario}
We will now present the results of our approach for multiuser scenario when multiple UTs are communicating simultaneously  with the LEO satellite constellation. This represents a more realistic scenario as compared to the single UT case. However, in multiuser scenario, the signal received at each satellite is a combination of data streams from all the UTs. The IUI degrades the uplink BER performance with the increase in number of UTs served, as depicted in Fig. \ref{fig:singlevsmulti}. During simulation, the distance between two UTs was fixed and two more UTs were uniformly placed in between the existing UTs (for four UTs scennario) to better visualize the effect of IUI on uplink communication performance.

\begin{figure}
    \centering
    \includegraphics[width=0.5\textwidth]{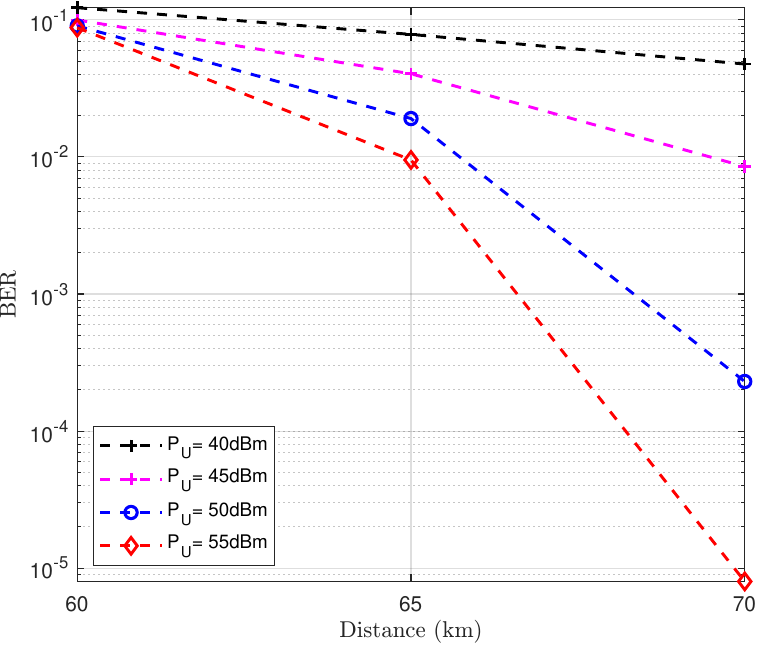}
    \caption{BER as a function of the UTs separation distance $D$ for different transmit power levels.}
    \label{fig:bervsdist}
\end{figure}

To observe the effect of separation distance between UTs, four UTs are placed at the corners of a  square centered around origin, with each side having a length of $D$ as illustrated in Fig. \ref{fig:const}. The separation between the UTs is then increased by expanding this square by increasing $D$. Figure \ref{fig:bervsdist} shows the uplink BER as a function of the UTs separation distance $D$ for different UTs' transmit power levels. As the separation $D$ increases, the BER improves due to reduced IUI. This effect is more evident at higher transmit powers, where the system is interference-limited, while at lower powers the noise dominance leads to flatter BER curves.

\section{Conclusion}
This paper proposed an IPAC framework for improved channel estimation in multi-LEO satellite multiuser networks based on dual-timescale Kalman filtering. By leveraging UTs’ position and velocity information obtained during the large-timescale downlink positioning phase, the framework enables improved channel estimation and cooperative data detection in the small-timescale uplink communications, reducing pilot overhead and improving spectral efficiency. Numerical results confirmed that the proposed approach achieves superior channel tracking accuracy and communication performance compared to the conventional position-information-agnostic and estimation-then-detection approaches. These results highlight the potential of IPAC as an enabler for future non-terrestrial 6G networks. Future work can exploit multiuser joint detection, instead of treating inter-user signals as interference at each satellite, to further improve the performance of the proposed approach in multiuser scenarios.

\bibliographystyle{IEEEtran}
\bibliography{references}

@article{andrews20246,
  title={{6G} Takes Shape},
  author={Andrews, Jeffrey G and Humphreys, Todd E and Ji, Tingfang},
  journal={IEEE BITS the Information Theory Magazine},
  year={2024},
  publisher={IEEE}
}

@article{liu2021leo,
  title={{LEO} satellite constellations for {5G} and beyond: How will they reshape vertical domains?},
  author={Liu et al., Shicong},
  journal=IEEE_M_COM,
  volume={59},
  year={2021},
  publisher={IEEE}
}

@article{ma2024integrated,
  title={Integrated Positioning and Communication via {LEO} Satellites: Opportunities and Challenges},
  author={Ma et al., Jie},
  journal={arXiv preprint arXiv:2411.14360},
  year={2024}
}

@article{zhang2024multi,
  title={Multi-satellite cooperative networks: Joint hybrid beamforming and user scheduling design},
  author={Zhang et al., Xuan},
  journal=IEEE_J_WCOM,
  volume={23},
  number={7},
  pages={7938--7952},
  year={2024},
  publisher={IEEE}
}

@article{arapoglou2010mimo,
  title={{MIMO} over satellite: A review},
  author={Arapoglou et al., Pantelis-Daniel},
  journal=IEEE_J_STCOM,
  volume={13},
  number={1},
  pages={27--51},
  year={2010},
  publisher={IEEE}
}

@ARTICLE{10750262,
  author={Darya, Abdollah Masoud and Abdallah, Saeed},
  journal=IEEE_J_COML, 
  title={Semi-Blind Channel Estimation for Massive {MIMO} {LEO} Satellite Communications}, 
  year={2025},
  volume={29},
  number={1},
  pages={75-79},
  doi={10.1109/LCOMM.2024.3495998}}

@article{li2023channel,
  title={Channel estimation for {LEO} satellite massive {MIMO} {OFDM} communications},
  author={Li, Ke-Xin and Gao, Xiqi and Xia, Xiang-Gen},
  journal=IEEE_J_WCOM,
  volume={22},
  number={11},
  pages={7537--7550},
  year={2023},
  publisher={IEEE}
}

@inproceedings{muns2019beam,
  title={Beam alignment and tracking for autonomous vehicular communication using IEEE 802.11 ad-based radar},
  author={Muns et al., Guillem Reus},
  booktitle={{INFOCOM} workshop},
  pages={535--540},
  year={2019},
  organization={IEEE}
}

@article{collaborative,
  title={Collaborative sensing in perceptive mobile networks: Opportunities and challenges},
  author={Xie, Lei and Song, Shenghui and Eldar, Yonina C and Letaief, Khaled B},
  journal=IEEE_M_WC,
  volume={30},
  number={1},
  pages={16--23},
  year={2023},
  publisher={IEEE}
}

@article{wei2022toward,
  title={Toward multi-functional {6G} wireless networks: Integrating sensing, communication, and security},
  author={Wei et al., Zhongxiang},
  journal=IEEE_M_COM,
  volume={60},
  number={4},
  pages={65--71},
  year={2022},
  publisher={IEEE}
}

@article{di2014location,
  title={Location-aware communications for {5G} networks: How location information can improve scalability, latency, and robustness of {5G}},
  author={Di Taranto et al., Rocco},
  journal=IEEE_M_SP,
  volume={31},
  number={6},
  pages={102--112},
  year={2014},
  publisher={IEEE}
}

@article{wymeersch20185g,
  title={{5G} mmWave positioning for vehicular networks},
  author={Wymeersch et al., Henk},
  journal=IEEE_M_WC,
  volume={24},
  number={6},
  pages={80--86},
  year={2018},
  publisher={IEEE}
}

@article{zhang2025positioning,
  title={Positioning-Aided Channel Estimation for Multi-{LEO} Satellite Cooperative Communications},
  author={Zhang et al., Yuchen},
  journal={arXiv preprint:2502.05808},
  year={2025}
}

@article{ferre2022leo,
  title={Is {LEO}-based positioning with mega-constellations the answer for future equal access localization?},
  author={Ferre et al., Ruben Morales},
  journal=IEEE_M_COM,
  volume={60},
  number={6},
  pages={40--46},
  year={2022}
}

@article{zheng2024leo,
  title={{LEO}- and {RIS}-empowered user tracking: A Riemannian manifold approach},
  author={Zheng, Pinjun and Liu, Xing and Al-Naffouri, Tareq Y},
  journal=IEEE_J_JSAC,
  year={2024},
  publisher={IEEE}
}

@article{you2024integrated,
  title={Integrated communications and localization for massive {MIMO} {LEO} satellite systems},
  author={You et al., Li},
  journal=IEEE_J_WCOM,
  year={2024},
  publisher={IEEE}
}

@inproceedings{7,
  title={Estimation of {CSI} for {mMIMO} based on received data using Kalman filter},
  author={Almamori, Aqiel and Mohan, Seshadri},
  booktitle={{CCWC}},
  pages={665--669},
  year={2018},
  organization={IEEE}
}

@article{8,
  title={Sparse doubly-selective channel estimation techniques for {OSTBC} {MIMO-OFDM} systems: A hierarchical Bayesian Kalman filter based approach},
  author={Srivastava et al., Suraj},
  journal=IEEE_J_COM,
  volume={68},
  number={8},
  pages={4844--4858},
  year={2020},
  publisher={IEEE}
}

@ARTICLE{9,
  author={Tang, Ruiguang and Zhou, Xiao and Wang, Chengyou},
  journal={IEEE Access}, 
  title={Kalman Filter Channel Estimation in 2 × 2 and 4 × 4 {STBC} {MIMO-OFDM} Systems}, 
  year={2020},
  volume={8},
  number={},
  pages={189089-189105},
  doi={10.1109/ACCESS.2020.3027377}}

@article{10,
  title={Adaptive tracking for beam alignment between ship-borne digital phased-array antenna and {LEO} satellite},
  author={Chen, Qin and Xu, Yuying and Song, Chunyi and Xu, Zhiwei},
  journal={J. Commun. \& Info. Networks},
  volume={4},
  number={3},
  pages={60--70},
  year={2019},
  publisher={PTP}
}

@ARTICLE{11,
  author={Yue, Tong and Liu, Aijun and Liang, Xiaohu},
  journal=IEEE_J_COML, 
  title={Block-Based Kalman Channel Tracking for {LEO} Satellite Communication With Massive {MIMO}}, 
  year={2023},
  volume={27},
  number={2},
  pages={645-649},
  doi={10.1109/LCOMM.2022.3224910}}

@article{12,
  title={Massive {MIMO} transmission for {LEO} satellite communications},
  author={You et al., Li},
  journal=IEEE_J_JSAC,
  volume={38},
  number={8},
  pages={1851--1865},
  year={2020},
  publisher={IEEE}
}

@article{13,
  title={Performance analysis of satellite communication system under the shadowed-Rician fading: A stochastic geometry approach},
  author={Jung, Dong-Hyun and Ryu, Joon-Gyu and Byun, Woo-Jin and Choi, Junil},
  journal=IEEE_J_COM,
  volume={70},
  number={4},
  pages={2707--2721},
  year={2022},
  publisher={IEEE}
}

@article{14,
  title={Downlink analysis and evaluation of multi-beam {LEO} satellite communication in shadowed Rician channels},
  author={Kim, Eunsun and Roberts, Ian P and Andrews, Jeffrey G},
  journal=IEEE_J_VT,
  volume={73},
  number={2},
  pages={2061--2075},
  year={2023},
  publisher={IEEE}
}

@article{15,
  title={Three-dimension massive {MIMO} for air-to-ground transmission: Location-assisted precoding and impact of {AoD} uncertainty},
  author={Xu, Youyun and Xia, Xiaochen and Xu, Kui and Wang, Yurong},
  journal={IEEE Access},
  volume={5},
  pages={15582--15596},
  year={2017},
  publisher={IEEE}
}

@article{16,
  title={Study on New Radio {(NR)} to support non-terrestrial networks},
  author={{3GPP TR 38.811 V15.4.0}},
  journal={Tech. Rep.},
  year={Sept. 2020}
}

@inproceedings{17,
  title={A {5G-NR} satellite extension for the QuaDRiGa channel model},
  author={Jaeckel, Stephan and Raschkowski, Leszek and Thieley, Lars},
  booktitle={EuCNC/{6G} Summit},
  pages={142--147},
  year={2022},
  organization={IEEE}
}

@article{18,
  title={Attenuation by atmospheric gases and related effects},
  author={{International Telecommunication Union (ITU)}},
  journal={Recommendation ITU-R},
  volume={25},
  pages={676--12},
  year={2019}
}

@inproceedings{19,
  title={The unscented Kalman filter for nonlinear estimation},
  author={Wan, Eric A and Van Der Merwe, Rudolph},
  booktitle={IEEE SSPCC Symp.},
  pages={153--158},
  year={2000}
}

@book{20,
  title={Inference and Learning from Data: Learning},
  author={Sayed, Ali H},
  volume={3},
  year={2022},
  publisher={Cambridge University Press}
}

@book{kay1993fundamentals,
  title={Fundamentals of statistical signal processing: estimation theory},
  author={Kay, Steven M},
  year={1993},
  publisher={Prentice-Hall, Inc.}
}

@article{fbkf,
  title={A forward-backward Kalman filter-based {STBC} {MIMO} {OFDM} receiver},
  author={Quadeer, Ahmed Abdul and Al-Naffouri, TY},
  journal={{EURASIP}},
  volume={2008},
  pages={158037},
  year={2008},
  publisher={Springer}
}

@article{jakes,
  title={Jakes fading model revisited},
  author={Dent, Paul and Bottomley, Gregory E and Croft, T},
  journal={Electronics letters},
  volume={29},
  number={13},
  pages={1162--1163},
  year={1993},
  publisher={IET}
}

@article{bello1963characterization,
  title={Characterization of randomly time-variant linear channels},
  author={Bello, Philip},
  journal={IEEE Trans. Commun.},
  volume={11},
  number={4},
  pages={360--393},
  year={1963},
  publisher={IEEE}
}

@inbook{bjornson2024introduction,
  title     = {Introduction to multiple antenna Commun. and reconfigurable surfaces},
  author    = {Bj{\"o}rnson, Emil and Demir, {\"O}zlem Tu{\u{g}}fe and others},
  chapter   = {7}, 
  publisher = {Now Publishers},
  year      = {2024}
}

@inproceedings{quadriga,
  title={QuaDRiGa: A {MIMO} channel model for land mobile satellite},
  author={Burkhardt, Frank and Jaeckel, Stephan and Eberlein, Ernst and Prieto-Cerdeira, Roberto},
  booktitle={{EuCAP}},
  pages={1274--1278},
  year={2014},
  organization={IEEE}
}

@article{hourani2024atmospheric,
  title={Atmospheric absorption loss for satellite communications},
  author={Hourani, A},
  journal={MATLAB Central File Exchange, Retrieved Feb},
  volume={28},
  year={2024}
}

@article{kassas2024ad,
  title={Ad astra: Simultaneous tracking and navigation with megaconstellation {LEO} satellites},
  author={Kassas, Zaher M and Khairallah, Nadim and Kozhaya, Sharbel},
  journal=IEEE_M_AES,
  volume={39},
  number={9},
  pages={46--71},
  year={2024},
  publisher={IEEE}}

@misc{celestrak2024tle,
  author       = {{CelesTrak}},
  title        = {{NORAD} Two-Line Element Sets},
  year         = {2024},
  howpublished = {\url{https://www.celestrak.com/NORAD/elements/}},
  note         = {[Online; accessed 28-Aug-2025]}
}

@article{heo2023mimo,
  title={{MIMO} satellite communication systems: A survey from the PHY layer perspective},
  author={Heo et al., Jehyun},
  journal=IEEE_J_STCOM,
  volume={25},
  number={3},
  pages={1543--1570},
  year={2023},
  publisher={IEEE}
}

@ARTICLE{9939157,
  author={Abdelsadek, Mohammed Y. and Kurt, Gunes Karabulut and Yanikomeroglu, Halim},
  journal={IEEE Open J. Commun. Soc.}, 
  title={Distributed Massive {MIMO} for {LEO} Satellite Networks}, 
  year={2022},
  volume={3},
  number={},
  pages={2162-2177},
  doi={10.1109/OJCOMS.2022.3219419}}

@ARTICLE{10851844,
  author={Wang et al., Qi},
  journal=IEEE_J_WCOM, 
  title={Multiple-Satellite Cooperative Information Communication and Location Sensing in {LEO} Satellite Constellations}, 
  year={2025},
  volume={24},
  number={4},
  pages={3346-3361},
  doi={10.1109/TWC.2025.3530083}}

@article{zhang2025cooperative,
  title={Cooperative sensing performance of multi-satellite uplink systems},
  author={Zhang et al., Xiaorui},
  journal={Physical Commun.},
  volume={71},
  pages={102678},
  year={2025},
  publisher={Elsevier}
}

@inproceedings{lin2024joint,
  title={Joint Doppler and Angle Tracking for {LEO} Satellite Communications Exploiting Orbit Characteristics},
  author={Lin, Chenlan and Chen, Xiaoming and Zhang, Zhaoyang},
  booktitle={{WCSP}},
  pages={993--998},
  year={2024},
  organization={IEEE}
}

@article{liu2022robust,
  title={Robust energy-efficient hybrid beamforming design for {mMIMO LEO} satellite Commun. systems},
  author={Liu et al., Yang},
  journal={IEEE Access},
  year={2022},
  publisher={IEEE}
}

@ARTICLE{930095,
  author={Papathanassiou, A. and Salkintzis, A.K. and Mathiopoulos, P.T.},
  journal={IEEE Personal Commun.}, 
  title={A comparison study of the uplink performance of {W-CDMA} and {OFDM} for mobile multimedia communications via {LEO} satellites}, 
  year={2001},
  volume={8},
  number={3},
  pages={35-43},
  doi={10.1109/98.930095}}

@article{arti2015channel,
  title={Channel estimation and detection in hybrid satellite--terrestrial communication systems},
  author={Arti, MK},
  journal=IEEE_J_VT,
  volume={65},
  number={7},
  pages={5764--5771},
  year={2015},
  publisher={IEEE}
}

@article{arti2016channel,
  title={Channel estimation and detection in satellite communication systems},
  author={Arti, MK},
  journal=IEEE_J_VT,
  volume={65},
  number={12},
  pages={10173--10179},
  year={2016},
  publisher={IEEE}
}

@article{gappmair2014new,
  title={New results on location-aware channel estimation for multibeam satellite links},
  author={Gappmair, Wilfried and Bergmann, Michael and Suesser-Rechberger, Barbara},
  journal=IEEE_J_COML,
  volume={18},
  number={8},
  pages={1355--1358},
  year={2014},
  publisher={IEEE}
}

\end{document}